\documentclass[10pt]{article} \ifdefined\anonymoussubmission
  \usepackage{tmlr}
\else
  \usepackage[accepted]{tmlr}
\fi

\usepackage{amsmath,amsfonts,bm}

\def\eqref#1{equation~\ref{#1}}

\def\1{\bm{1}}

\DeclareMathAlphabet{\mathsfit}{\encodingdefault}{\sfdefault}{m}{sl}
\SetMathAlphabet{\mathsfit}{bold}{\encodingdefault}{\sfdefault}{bx}{n}

\usepackage{hyperref}
\usepackage{url}

\usepackage[T1]{fontenc}
\usepackage{graphicx}
\usepackage{booktabs}       \usepackage{amsfonts}       \usepackage{nicefrac}       \usepackage{microtype}      \usepackage{xcolor}         \usepackage{todonotes}
\usepackage{amsmath, amssymb}
\usepackage{subcaption}
\usepackage{adjustbox}
\usepackage{multirow}
\usepackage{wrapfig}
\usepackage{float}
\usepackage{needspace}
\usepackage{placeins}
\usepackage{array}
\usepackage{tabularx}
\usepackage{tikz}
\usepackage{tcolorbox}
\usepackage{colortbl}
\usepackage{soul}
\usepackage{multicol}
\usepackage{titletoc}
\usepackage{xspace}

\usepackage[capitalize]{cleveref}
\crefname{section}{Sec.}{Secs.}
\Crefname{section}{Section}{Sections}
\crefname{table}{Tab.}{Tabs.}
\Crefname{table}{Table}{Tables}

\def\eg{\emph{e.g.}\xspace} 
\def\ie{\emph{i.e.}\xspace}

\newcommand{\acronym}{\textsc{Squeeze3D}}
\newcommand{\best}[1]{\textbf{#1}}
\newcommand{\secondbest}[1]{\underline{#1}}

\definecolor{tabfirst}{rgb}{1, 0.7, 0.7} \definecolor{tabsecond}{rgb}{1, 0.85, 0.7} \definecolor{tabthird}{rgb}{1, 1, 0.7} \definecolor{projecturlblue}{RGB}{70,150,220}

\definecolor{revisionpink}{RGB}{214,51,132}
\disablepinktrue
\ifdisablepink
  \newcommand{\pink}[1]{#1}
  \newcommand{\pinkall}{}
\else
  \newcommand{\pink}[1]{{\color{revisionpink}#1}}
  \newcommand{\pinkall}{\color{revisionpink}}
\fi

\def\month{09}
\def\year{2026}
\def\openreview{\url{https://openreview.net/forum?id=XXYGHfqzvA}}

\title{\acronym: Extreme Neural Compression with Latent Space Bridging}

\ifdefined\anonymoussubmission
\author{\name Anonymous Author(s) \email anon@example.com \\
\addr Anonymous Institution}
\else
\author{\name Rishit Dagli \email rishit@cs.toronto.edu \\
\addr University of Toronto and Vector Institute
\AND
\name Yushi Guan \email guanyushi@cs.toronto.edu \\
\addr University of Toronto and Vector Institute
\AND
\name Sankeerth Durvasula \email sankeerth@cs.toronto.edu \\
\addr University of Toronto and Vector Institute
\AND
\name Mohammadreza Mofayezi \email mofayezi@cs.toronto.edu \\
\addr University of Toronto and Vector Institute
\AND
\name Nandita Vijaykumar \email nandita@cs.toronto.edu \\
\addr University of Toronto and Vector Institute
\AND
\name \href{https://squeeze3d.github.io}{\texttt{\textcolor{projecturlblue}{https://squeeze3d.github.io}}}
}
\fi

\begin{document}

\maketitle

  \vspace{-0.6em}
  \begin{center}
    \vspace{-0.5em}
    \includegraphics[width=\linewidth]{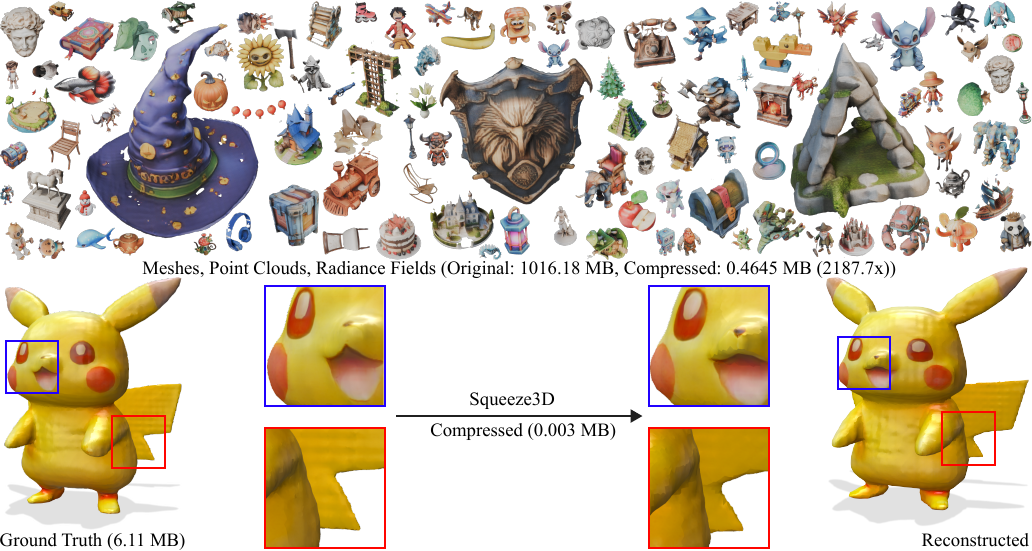}\captionof{figure}{\textbf{\acronym} allows for extreme compression of 3D models while preserving perceptual quality. \textit{Top:} Our method compresses a diverse collection of 3D models. \textit{ Bottom:} Comparison between the original model (6.11 MB) and the reconstruction after compression (0.003 MB).}
    \label{fig:teaser}
  \end{center}
  \vspace{0.5em}

\begin{abstract}
We propose \acronym, a novel framework that leverages implicit prior knowledge learnt by existing pre-trained encoders and decoders to compress 3D data at extremely high compression ratios. Our approach bridges the latent spaces between a pre-trained encoder and a pre-trained decoder model through trainable mapping networks. Any 3D asset represented as a mesh, point cloud, or radiance field is first encoded by the pre-trained encoder and then transformed (\ie \emph{compressed}) into a highly compact latent code by a mapping network. This latent code can effectively be used as an extremely compressed representation of the mesh, point cloud, or radiance field.
A mapping network transforms the compressed latent code into the latent space of a powerful generative model; the decoder of this generative model then recreates the original 3D asset (\ie \emph{decompression}).
\acronym\ is trained entirely on generated synthetic data and does not require any 3D datasets. The \acronym\ architecture can be flexibly used with existing pre-trained 3D encoders and existing generative models. It can flexibly support different formats, including meshes, point clouds, and radiance fields.
Our experiments demonstrate that \acronym\ achieves compression ratios of up to 2187$\times$ for textured meshes, 58.5$\times$ for point clouds, and more than 650$\times$ for radiance fields while maintaining visual quality comparable to many existing methods. \acronym\ only incurs a small compression and decompression latency since it does not involve training object-specific networks to compress an object.
\ifdefined\anonymoussubmission
We provide extensive additional visual results at \url{https://anonymous.4open.science/w/tmlr-s3d-website/} and code at \url{https://anonymous.4open.science/r/tmlr-s3d-code}.
\fi
\end{abstract}

\section{Introduction}
\label{sec:introduction}

The rapid advancement of 3D data acquisition and representation technologies over the past decade has significantly expanded the availability and generation of high-resolution 3D content across various domains in different formats, including meshes, point clouds, and radiance fields (which could be extracted from a NeRF~\citep{nerfs} or a 3DGS~\citep{kerbl3Dgaussians}). The widespread use of 3D data necessitates the development of techniques that enable efficient transmission, storage, and processing of large-scale 3D representations. To this end, \emph{compression} and the use of \emph{compressed representations} for 3D data are of utmost importance, e.g., in streaming, autonomous navigation, digital twins, remote sensing.

A large body of work proposes techniques to compress meshes, point clouds, neural radiance fields (NeRFs)~\citep{nerfs}, and 3D Gaussian Splats (3DGS)~\citep{kerbl3Dgaussians}. These approaches aim to maximize the compression ratio while retaining reconstruction quality. For example, traditional mesh decimation techniques~\citep{10.1145/142920.134010, 10.1145/258734.258849, 10.1007/978-3-319-11740-9_25, https://doi.org/10.1111/cgf.13932} remain foundational for polygon reduction, but their reliance on handcrafted simplification rules limits their ability to preserve fine geometric details at extreme compression ratios. MPEG's G-PCC and V-PCC standards~\citep{schwarz2018emerging, liu2019comprehensive} use projection-based methods for point cloud compression; however, these approaches incur overheads for representing fine details and packing. Prior works also propose a range of compression methods for NeRFs~\citep{10656595, 10.1145/3641519.3657399, 10.1145/3528233.3530727, M_ller_2022} or 3DGS models~\citep{fan2024lightgaussianunbounded3dgaussian, Lee_2024_CVPR}. Several works propose autoencoder-style networks that compress 3D models into small latent vectors~\citep{10.1007/978-3-031-72649-1_16, Hahner_2022_WACV, NEURIPS2020_68dd09b9, chen2025deepcompressionautoencoderefficient}.
The maximum compression ratios achieved by these methods are of the order of 100$\times$ for meshes and of the order of 10$\times$ for point clouds, but typically much lower.

Our goal is to develop a framework for \emph{extreme} compression of 3D data stored in any format while retaining high visual quality. We emphasize that \acronym\ is designed to compress \emph{arbitrary, real-world} 3D assets (e.g., artist-created meshes and 3D scans). Such extreme ratios are most valuable at scale: streaming or archiving collections of millions of 3D assets, delivering content to bandwidth-constrained or on-device clients, and settings where the decoder is a shared, one-time asset that is amortized across all compressed objects (\Cref{sec:amortization}). Recent years have seen significant and continued advances and development of powerful generative models. In this work, we aim to leverage the implicit prior knowledge learnt by the decoders of powerful 3D generative models~\citep{xu2024instantmesh, jun2023shapegeneratingconditional3d} to enable extreme compression ratios. Recent works also propose techniques to leverage generative models for 3D compression~\citep{zhang20243drepresentation512bytevariationaltokenizer, cui2024neusdfusionspatialawaregenerativemodel, 10.1007/978-3-031-72649-1_16, NIPS2017_7a98af17, zhang2024g3pt, chang20253dshapetokenizationlatent, roessle2024l3dg}, and one achieves extreme compression for meshes~\citep{zhang20243drepresentation512bytevariationaltokenizer}. However, these approaches require training specialized encoders and generative models for a single 3D format. In contrast, our goal is to flexibly use \emph{existing} encoders and decoder models that provide adaptability as encoders/generative models evolve and flexibility across 3D formats.

We propose \acronym, a compression framework that generates a highly compressed latent vector that can be used to recreate the original 3D data using an existing pre-trained decoder model. \acronym\ comprises three key components: (1) the input 3D data is encoded with a pre-trained encoder. This allows us to extend \acronym\ to other 3D encoders. (2) We train two small neural networks that we call the \emph{forward mapping network} and the \emph{reverse mapping network}. The forward mapping network maps the encoded representations into an extremely compressed latent space. The reverse mapping network converts the code from the compressed latent space to the latent space of the generative model. (3) We use a pre-trained decoder of a generation model to generate the original 3D data using the code generated by the reverse mapping network.
\acronym\ can be flexibly implemented with any pre-trained encoder and decoder of a generative model.

The forward and reverse mapping networks are trained for any given encoder-generator pair. We first synthetically generate a 3D dataset via random prompts to the generator model. This 3D dataset is encoded using the pre-trained encoder. The set of latents produced from the pre-trained encoders (training inputs) and their corresponding latents from the pre-trained generator (ground truth) are used to train the forward and reverse mapping networks. We propose a loss function that minimizes redundant information in the compressed latent space.

\acronym\ can be flexibly applied to 3D data in different formats. We implement and evaluate our method for mesh, point cloud, and radiance field compression using 3 existing encoders and 5 existing generative models. We demonstrate that our method achieves compression ratios significantly higher than or on par with those of existing compression techniques for meshes, point clouds, and radiance fields, with reconstruction quality that is on par with many prior approaches. We demonstrate a compression ratio of 2187$\times$ for a subset of the Objaverse~\citep{deitke2023objaversexluniverse10m3d} dataset, 58.5$\times$ on a subset of ShapeNet~\citep{chang2015shapenetinformationrich3dmodel}, and more than 650$\times$ across our radiance-field evaluations~\citep{irshad2024nerf}. While \acronym\ expectedly cannot achieve state-of-the-art reconstruction quality, we qualitatively show that it is able to retain high visual quality in many settings.

\paragraph{Contributions.} (1) To the best of our knowledge, this is the first framework that leverages \emph{pre-existing, pre-trained} generative-model decoders to enable extreme compression of 3D data.

(2) We demonstrate the feasibility of aligning different latent spaces originating from neural networks with different structures, optimization objectives, and training distributions. To make this bridging robust at extreme compression, we introduce a \emph{Gram loss} that prevents the compressed latent from collapsing onto a few dominant directions (which we show is essential in our ablations), together with a synthetic data-generation recipe that trains the mapping networks without requiring any 3D datasets.

(3) We evaluate \acronym\ for mesh, point cloud, and radiance field compression and demonstrate that generative models are a promising approach for extreme compression of 3D models. \acronym\ can be flexibly extended to different encoders, generative models, and 3D formats.

\section{Related Works}
\label{sec:rw}

\subsection{Neural Graphic Primitives}
\label{sec:ngp}

Neural networks are increasingly being used to represent 2D images~\citep{NEURIPS2020_53c04118, NEURIPS2020_55053683, gao2024hsirenimprovingimplicitneural}, 3D objects and scenes~\citep{sitzmann2019scene, jiang2020local, peng2020convolutional, NEURIPS2020_55053683, https://doi.org/10.1111/cgf.14505, davies2021effectivenessweightencodedneuralimplicit, nerfs, martel2021acornadaptivecoordinatenetworks}, surface representations~\citep{takikawa2021nglod, tang2022nerf2mesh, rakotosaona2023nerfmeshingdistillingneuralradiance, 10.1145/3641519.3657399}, occupancy networks~\citep{mescheder2019occupancy, chen2019learning}, and signed distance fields~\citep{michalkiewicz2019implicit, park2019deepsdf, atzmon2020sal}. These methods, out of the box, can also be used to compress 3D models in some format since the learned neural network weights are often already significantly smaller.
Many NGP compression methods often employ standard neural network techniques to compress MLPs by knowledge distillation~\citep{shi2022distilled}, pruning~\citep{isik2021neural, xie2023hollownerf, jung2024prunerf}, quantization~\citep{shenaccelerating, zhong2023vq, yang2023vq, gordon2023quantizing, zhang2024rate}, factorizing tensor grids~\citep{gao2023strivecsparsetrivectorradiance, obukhov2023ttnf}, low-rank approximation~\citep{shi2022distilled, app142311277, NEURIPS2022_5ed5c3c8, ji2025efficientlearningsineactivatedlowrank}, and using codebooks~\citep{li2023compact, 10204677} for quantization. Another approach is to compress feature grids or learnable embeddings~\citep{shin2024binary, Chen_2024_CVPR, pham2024neural} or extracted voxels~\citep{9968173, chen2022tensorf}, rather than compressing the MLP. Another set of approaches combines many of these orthogonal improvements to compressing NeRFs~\citep{10656595}. However, these methods often require training networks per scene or object, incurring significant compression latencies.

\subsection{3D Autoencoders and Generators}
\label{sec:rw:autoenc}

Autoencoders encode inputs into a latent space and then decode them back into the original input space using an encoder-decoder pair; thus, they could be used as compression algorithms.
Early approaches relied on volumetric representations, discretizing 3D shapes into voxel grids to leverage ConvNets for encoding and decoding~\citep{9290656, Hess_2023, li2023voxformersparsevoxeltransformer, brock2016generativediscriminativevoxelmodeling, molnar2024variational}. While these approaches are effective for regular grid structures, these methods faced scalability challenges due to cubic memory growth with resolution increases.
Subsequent advancements focused on spectral methods for encoding 3D shapes in frequency domains, offering compact latent representations but requiring precomputed basis functions that limited generalizability across shape categories~\citep{zamorski2020adversarial, park2019deepsdf}.

Several works~\citep{zhao2023michelangelo, chen20243dtopia, wu2024direct3d, lan2024ln3diff} use a VAE~\citep{kingma2013auto} to compress 3D data into a compact latent which could be used as a compressed representation. There also exist many approaches that train generators to reconstruct radiance fields~\citep{kosiorek2021nerf, irshad2024nerf, xiang2024structured}, Gaussians~\citep{wewer2024latentsplat, ma2024shapesplat, xiang2024structured}, or voxels~\citep{ren2024xcube, xiang2024structured}. Some recent works also pose the problem as learning in a token-space~\citep{chang20253dshapetokenizationlatent, chen2024meshanythingartistcreatedmeshgeneration, Siddiqui_2024_CVPR, zhang20243drepresentation512bytevariationaltokenizer, 10.1145/3552457.3555731, 10849903, DBLP:conf/dicta/LuoJP23, 10.1145/3512527.3531423, huo2025renderingoriented3dpointcloud, 9922700, Cui_Long_Feng_Li_Kai_2023}, or with diffusion models~\citep{10.1007/978-3-031-72649-1_16, anonymous2024diffpcc, roessle2024l3dg, galvis2024scdiff3dshapecompletion, rs17071215, 10518035}.
One such generative model~\citep{zhang20243drepresentation512bytevariationaltokenizer} is able to compress meshes with high compression ratios. A rapidly growing line of recent shape VAEs and native 3D generators learns increasingly compact latents with high-quality reconstruction~\citep{zhang20233dshape2vecset, li2024craftsman, chen2025dora, cho2025latent64, li2025triposg, xiang2025trellis2, chen20243dtopia}. Auto-decoder approaches such as ROAD~\citep{road2023} likewise encode shapes into compact codes, but require a per-shape optimization at encoding time. Compared to these methods that use autoencoders or generative models for compression, \acronym\ does not require training specialized encoder-generators for each representation, nor any per-shape optimization at encode time. Instead, \acronym\ aims to use existing encoders and generative models, and is in fact complementary to this line of work: any of these stronger generators can be adopted as the decoder within our framework. This enables flexibly adapting the approach as encoders and generative models evolve and supporting different 3D formats.

The work closest to our method is Generative Latent Coding (GLC)~\citep{10658512}, which trains an autoencoder-style generative model to compress images, particularly compressing the latent representations for an image obtained through VQ-VAE~\citep{NIPS2017_7a98af17}. However, this method is not designed for 3D data, requires access to an encoder and decoder trained jointly, and only tests one encoder-decoder pair.

\section{Method}
\label{sec:method}

\begin{figure*}[tb]
    \centering
    \includegraphics[width=\textwidth]{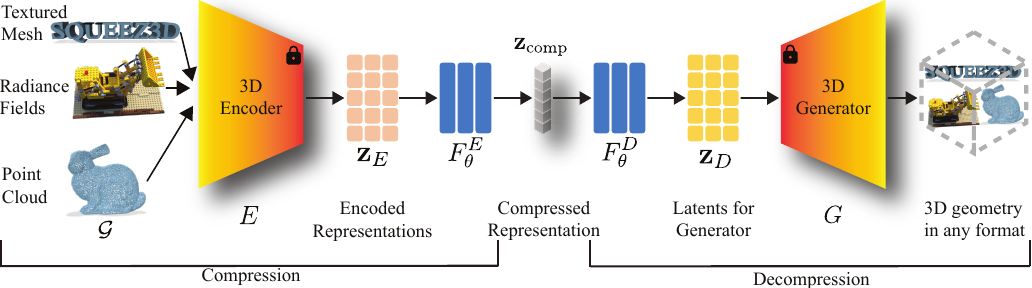}
    \caption{\textbf{Overview of our Method.} \acronym\ bridges arbitrary latent spaces between 3D encoders and generators through trainable mapping networks. During compression, a 3D geometry is encoded and then transformed into a compact representation via the forward mapping network. During decompression, the reverse mapping network converts this representation into the generator's latent space, which is then used to reconstruct the original geometry.}
    \label{fig:methods}
\end{figure*}

In this work, we introduce \acronym, a technique to generate highly compressed representations of 3D models by leveraging the implicit prior knowledge learnt by decoders of 3D generative models. We also leverage the availability of many 3D encoders to support an extensible set of 3D formats. 

The \acronym\ architecture is summarized in~\Cref{fig:methods}. The key ideas of \acronym\ are as follows. (1) We leverage existing 3D encoders to generate encoded representations for a given 3D format. Thus, the \acronym\ architecture can be extended to support different 3D formats by using new or existing encoders. This approach also enables smaller mapping networks, as we now introduce. (2) We use small neural networks to convert from this encoded representation to a highly compressed latent representation (during compression) and then back into the latent space of a 3D generative model (for decompression). We refer to these neural networks as the forward and reverse mapping networks, and they effectively map between the latent space of a 3D encoder and that of a 3D generative model. Thus \acronym\ can leverage a new 3D generative model by retraining the mapping networks. (3) We propose an additional loss term that enables robust training of the mapping networks to generate a highly compact latent representation that can be used to store/transmit the 3D model.
We share an overview of the notation we use to describe our method in~\Cref{tab:notation}.

\begin{table}[tb]
    \centering
    \caption{\textbf{Notation.} The notation we use to describe our method.}
    \small
    \resizebox{\textwidth}{!}{\begin{tabularx}{\linewidth}{lX|lX}
        \toprule
        \rowcolor{blue!15}Symb. & Description & Symb. & Description \\
        \midrule
        $\mathcal{G}$ & A 3D geometry in some format & $E$ & 3D encoder model: $E(\mathcal{G}) \mapsto \mathbf{z}_E \in \mathbb{R}^{d_E}$\\
        $G$ & 3D decoder model: $G(\mathbf{z}_G, c) \mapsto \mathcal{G}'$ & $\mathbf{z}_E$ & Latent code from the encoder: $\mathbf{z}_E \in \mathbb{R}^{d_E}$\\
        $\mathbf{z}_G$ & Latent code for the generator: $\mathbf{z}_G \in \mathbb{R}^{d_G}$ & $\mathbf{z}_{\text{comp}}$ & Compressed representation $\mathbf{z}_{\text{comp}} \in \mathbb{R}^{d_C}$\\
        $c$ & Conditioning information (\eg text prompt, image) for $G$ & $F_\theta^E$ & Forward mapping network: $F_\theta^E(\mathbf{z}_E) \mapsto \mathbf{z}_{\text{comp}}$\\
        $F_\theta^D$ & Reverse mapping network: $F_\theta^D(\mathbf{z}_{\text{comp}}) \mapsto \mathbf{z}_G$ & $d_C$ & Dimensionality of compressed representation\\
        $d_G$ & Dimensionality of generator latent space & $d_E$ & Dimensionality of encoder latent space\\
        \bottomrule
    \end{tabularx}}
    \label{tab:notation}
\end{table}
Mapping networks offer two major benefits over autoencoders and encoder-generator pairs, such as MeshAnything~\citep{chen2024meshanythingartistcreatedmeshgeneration}: (1) the mapping networks typically provide significantly higher compression ratios than existing VAE approaches; and (2) they provide more flexibility in the choice of 3D format. For instance, InstantMesh and LRM require multi-view images as input rather than a mesh, so they cannot be used directly as 3D-to-3D autoencoders, whereas the mapping-network design lets us pair them with any 3D encoder.

\subsection{Bridging Latent Spaces}
\label{sec:bridge}

\acronym\ comprises two pre-trained models: (1) a 3D encoder $E$ that maps 3D representations to a latent space, and (2) a 3D generator $G$ that synthesizes 3D models of the same initial representation. We use the term ``generator'' to refer to the model whose weights and learnt priors we leverage; at decompression time we only use its \emph{decoder} component, $G$, which maps a latent code to a 3D asset. For latent-diffusion generators such as LION~\citep{zeng2022lion}, this means we use the decoder that maps a latent to a point cloud and \emph{bypass} the diffusion sampler entirely. Consequently, decompression is a deterministic feed-forward pass: a given compressed code always reconstructs the same 3D asset. For a given 3D representation (mesh, point cloud, radiance field) $\mathcal{G}$, the pre-trained encoder $E$ produces a latent representation
\begin{equation}
    \mathbf{z}_E = E(\mathcal{G}) \in \mathbb{R}^{d_E}
\end{equation}
This latent $\mathbf{z}_E$ encapsulates $\mathcal{G}$, but is not in a highly compressed format. Additionally, we cannot directly use this representation with the generator $G$, as $G$ operates in a different latent space. The generator $G$ synthesizes a 3D model $\mathcal{G}'$ given a latent code $\mathbf{z}_G$ and in some cases conditioning information $c$, $\mathcal{G}' = G(\mathbf{z}_G, c)$. To bridge these disparate latent spaces, we train two mapping networks:

\textbf{Forward Mapping Network $F_\theta^E$}: Maps from the encoder's latent space to the compressed space, $\mathbf{z}_{\text{comp}} = F_\theta^E(\mathbf{z}_E)$.
\textbf{Reverse Mapping Network $F_\theta^D$}: Maps from the compressed space to the generator's latent space, $\mathbf{z}_G = F_\theta^D(\mathbf{z}_{\text{comp}})$.

We train $F_\theta^E$ and $F_\theta^D$ together and keep the encoder $E$ and decoder $G$ networks frozen.
\paragraph{Compression.} To compress any 3D model $\mathcal{G}$, the model is first encoded using the pre-trained encoder $E$ and then mapped into a highly compressed latent $\mathbf{z}_{\text{comp}}$ using the forward mapping network $F_\theta^E$
\begin{equation}
    \mathbf{z}_{\text{comp}} = F_\theta^E\left(E\left(\mathcal{G}\right)\right)
\end{equation}
\paragraph{Decompression.} To decompress the model from its highly compressed latent representation $\mathbf{z}_{\text{comp}}$, we use the reverse mapping network $F_\theta^D$ to obtain the latent in the 3D generator space. The 3D generator then reconstructs the original 3D model $\mathcal{G}$ as $\mathcal{G}'$
\begin{equation}
    \mathcal{G}' = G\left(F_\theta^D\left(\mathbf{z}_{\text{comp}}\right)\right)
\end{equation}
\pink{Each encoder-generator pair and each choice of $d_C$ requires a separately trained forward/reverse mapping pair. Once trained, that pair is reused for all objects.}

\subsection{Training \acronym}
\label{sec:improve}

\begin{figure*}[tb]
    \centering
    \includegraphics[width=\linewidth]{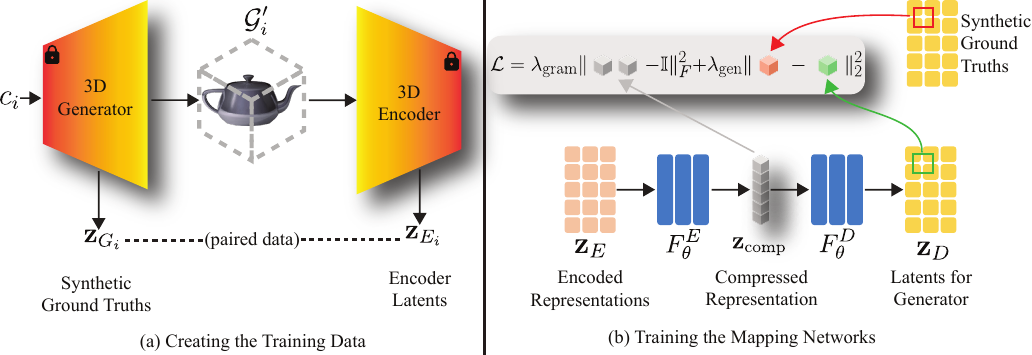}
    \caption{\textbf{Training \acronym.} We show an overview of (a) our process of creating synthetic data to train the mapping networks and (b) our process of training the mapping networks.}
    \label{fig:training}
\end{figure*}

In order to train the \acronym\ architecture, we need to train the mapping networks for any given pair of pre-trained 3D encoders and pre-trained 3D generator models. To train these mapping networks, we need training samples from the encoder's latent space and the corresponding latents in the generator's latent space. These latents in the generator's latent space serve as ``ground truth'' samples during the training process. We summarize our training process in~\Cref{fig:training}. We now describe how to generate a training dataset with samples from both of these latent spaces.

Given a pre-trained 3D generator $G'$ and encoder $E$, we first sample a diverse collection of conditioning inputs $\mathcal{C} = \{c_i\}_{i=1}^N$ appropriate for the generator model (\eg text prompts for text-to-3D generators, images for image-to-3D generators, or random noise for unconditional generators). For each conditioning input $c_i$, we sample a latent vector $\mathbf{z}_{G'_i}$ from the generator $G'$ and then synthesize a 3D model using the generator: $\mathcal{G}'_i = G(\mathbf{z}_{G'_i}, c_i)$. Then we encode this synthetic or generated 3D model using the pre-trained encoder $E$, $\mathbf{z}_{E_i} = E(\mathcal{G}'_i)$. This methodology gives us paired synthetic data, i.e., latents, for any given pair of encoder and generator: $\{(\mathbf{z}_{E_i}, \mathbf{z}_{G'_i})\}_{i=1}^N$, which provides the necessary supervision for training our mapping networks.

The mapping networks $F_\theta^E$ and $F_\theta^D$ are trained together on the generated dataset using the loss shown in~\Cref{eq:loss}.
\begin{equation}
    \mathcal{L} = \underbrace{\lambda_{\text{gram}}\|F_\theta^E(\mathbf{z}_E)F_\theta^E(\mathbf{z}_E)^\top - \mathbb{I}\|_F^2}_{\text{orthogonality of compressed representation}}
    + \underbrace{\lambda_{\text{gen}}\|F_\theta^D\left(F_\theta^E\left(\mathbf{z}_E\right)\right) - \text{GT}\|_2^2}_{\text{latent-space MSE}},
    \label{eq:loss}
\end{equation}
where GT represents the synthetic ground-truth latents and $\|\cdot\|_F$ denotes the Frobenius norm.

The loss function includes a \emph{latent-space MSE} term that minimizes the difference between the generated latents and the corresponding ground-truth latents. We also add another term, which we refer to as \emph{Gram loss}. When training \acronym\ with only the latent-space MSE term, we found that the mapping networks concentrate information along a small subset of dimensions, effectively rendering many dimensions redundant.

To understand this, we empirically analyzed the latent vectors $\mathbf{z}_{\text{comp}} = F_\theta^E(\mathbf{z}_E) \in \mathbb{R}^{d_C}$ produced by our forward mapping network. For any batch of size $B$ of encoded 3D models $\{\mathbf{z}_{E_i}\}_{i=1}^B$, we can compute the matrix $\mathbf{Z} \in \mathbb{R}^{B \times d_C}$ where each row is $F_\theta^E(\mathbf{z}_{E_i})$. First, we observe that if we do a singular value decomposition $\mathbf{Z} = \mathbf{U\Sigma V}^\top$, the singular values in $\Sigma=\operatorname{diag}(\sigma_1, \sigma_2, \ldots, \sigma_{d_C})$ exhibit an extremely skewed distribution: $\sigma_1 \gg \sigma_2 \gg \cdots \gg \sigma_{d_C}$ where $\sigma_i$ represents the $i$-th singular value of $\mathbf{Z}$, arranged in descending order. The condition number $\kappa = \frac{\sigma_1}{\sigma_{d_C}}$ is typically very large, indicating that the effective rank of $\mathbf{Z}$ is much lower than $d_C$.

Second, we observe that the correlation matrix $\mathbf{C} = \frac{1}{B}\mathbf{Z}^\top\mathbf{Z} \in \mathbb{R}^{d_C \times d_C}$ has many off-diagonal elements with large magnitudes in comparison with diagonal elements. This indicates that the dimensions of the compressed representation encode redundant information. Particularly,
\begin{equation}
    d_{\text{eff}} = \frac{(\sum_{i=1}^{d_C} \lambda_i)^2}{\sum_{i=1}^{d_C} \lambda_i^2} \ll d_C,
\end{equation}
where $\lambda_i$ are the eigenvalues. These observations indicate that most of the information in the compressed representation was concentrated along a few dominant directions, with most dimensions contributing negligibly. For compression, this represents a severe inefficiency in utilizing the available parameter budget.

To address this, we propose the Gram loss term that is computed on the outputs of the first mapping network $F_\theta^E$ to force the outputs of $F_\theta^E$ to be orthonormal. A semi-orthogonal matrix $\mathbf{A} \in \mathbb{R}^{m \times n}$ is defined as a matrix that satisfies either $\mathbf{A}\mathbf{A}^\top = \mathbf{I}_m$ (if $m \leq n$) or $\mathbf{A}^\top\mathbf{A} = \mathbf{I}_n$ (if $n \leq m$). Our Gram loss term when minimized forces $F_\theta^E(\mathbf{z}_E)F_\theta^E(\mathbf{z}_E)^\top \approx \mathbb{I}$, which is precisely the condition for $F_\theta^E(\mathbf{z}_E)$ to be a semi-orthogonal matrix (when $d_C \leq d_E$).

\section{Experiments}
\label{sec:exp}

\begin{figure*}[tb]
    \centering
    \includegraphics[width=\linewidth]{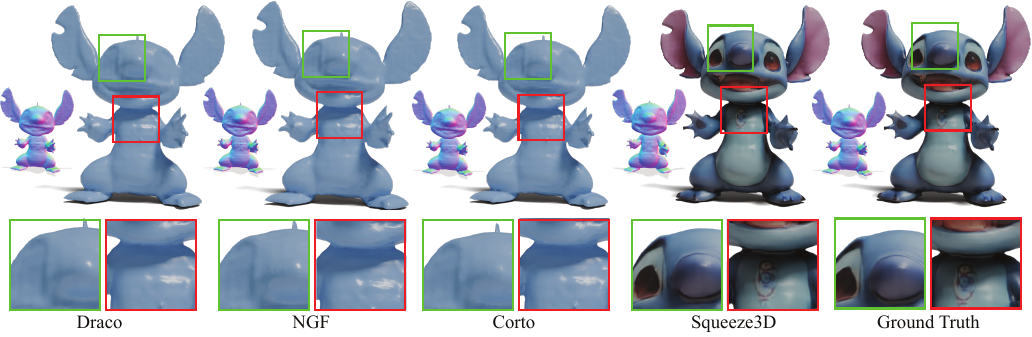}
    \caption{\textbf{Qualitative mesh compression results.} We compare \acronym\ to state-of-the-art methods. Our approach maintains visually important geometric details. Additional qualitative comparisons are provided in Appendix~\ref{app:qualmesh}.}
    \label{fig:results}
\end{figure*}

\subsection{Experimental Setup}
\label{app:experimental}

We train \acronym\ to compress three 3D formats: textured 3D meshes, point clouds, and radiance fields, \emph{i.e.}, grids of (rgb$\sigma$). For \emph{3D meshes}, we train our approach with MeshAnything~\citep{chen2024meshanythingartistcreatedmeshgeneration} as the encoder and train mapping networks for three 3D generators: Shap-E~\citep{jun2023shapegeneratingconditional3d}, OpenLRM~\citep{hong2024lrmlargereconstructionmodel, openlrm}, and InstantMesh~\citep{xu2024instantmesh}. For \emph{point clouds}, we train mapping networks for PointNet++~\citep{NIPS2017_d8bf84be} as the encoder and LION~\citep{zeng2022lion} as the decoder. For \emph{radiance fields}, we train mapping networks for NeRF-MAE~\citep{irshad2024nerf} as both the encoder and the generator. We present additional implementation details in~\Cref{app:additional} and the mapping-network architecture in~\Cref{app:networkdesign}.

\paragraph{Training Dataset Creation.} We use the method described in~\Cref{sec:improve} to train \acronym\ for each of our evaluated encoder-generator pairs. We now list the datasets that were used to create these latent training datasets.

\textbf{Shap-E~\citep{jun2023shapegeneratingconditional3d} as the generator.} We build a list of 2500 prompts using LLaMA3, each of which is used four times to build a dataset of 10,000 objects.

\textbf{LRM~\citep{hong2024lrmlargereconstructionmodel, openlrm} or InstantMesh~\citep{xu2024instantmesh} as the generator.} We rendered 10,000 random objects from Objaverse~\citep{deitke2023objaversexluniverse10m3d} which serve as image conditions. We also make sure that our rendering follows any conventions the 3D generator expects the input images to follow, for instance, white backgrounds or no backgrounds.

\textbf{LION~\citep{zeng2022lion} as the generator.} We were able to generate plausible point clouds of 3D objects without any conditioning data from random noise.

\textbf{NeRF-MAE~\citep{irshad2024nerf} as the generator.} We generate radiance fields from the NeRF-MAE~\citep{irshad2024nerf} dataset.

\pink{While many other compression methods require the 3D objects to have certain properties like watertightness, we do not impose any such constraints. We emphasize that, although the mapping networks are \emph{trained} on synthetic latents produced by the generator, they are \emph{evaluated} entirely on standard, third-party 3D data that was not produced by our pipeline: artist-created assets from Objaverse~\citep{deitke2023objaversexluniverse10m3d} and ShapeNet~\citep{chang2015shapenetinformationrich3dmodel}, real-world scanned objects from Amazon Berkeley Objects (ABO) (Appendix~\ref{sec:collection}), and captured scenes from NeRF-MAE~\citep{irshad2024nerf}. Our reported numbers therefore measure out-of-distribution generalization to real, human-made, and scanned 3D content rather than reconstruction of the generator's own outputs. We split the datasets into training (80\%) and validation (10\%) sets generated from our approach. We also built a test (10\%) set ($\mathcal{T}$-set) of objects from the datasets: Objaverse, ShapeNet, and NeRF-MAE, on which we report our metrics. Objaverse does not have pre-specified train and test splits. While our train and test splits are specified based on the splits used by MeshAnything~\citep{chen2024meshanythingartistcreatedmeshgeneration}, we do not have access to the splits used for training the generation models. This could cause the test set we build to be a part of the training set of the generation models we use. Thus, we separately construct another test set ($\mathcal{O}$-set) of 500 3D meshes (see~\Cref{sec:collection}) which are not in the Objaverse dataset, and report results on this test set.}

\paragraph{Evaluation metrics.} We report the standard widely used metrics, PSNR $\uparrow$, MS-SSIM $\uparrow$, and LPIPS $\downarrow$~\citep{8578166} for reconstruction quality of meshes and radiance fields. We report standard metrics, PCQM $\uparrow$~\citep{9123147}, and PointSSIM $\uparrow$~\citep{9106005} for reconstruction quality of point clouds. We give the precise definitions of all metrics and our rendering protocol in~\Cref{app:metrics}. We additionally report 3D geometric distortion metrics, Chamfer Distance $\downarrow$ and EMD $\downarrow$, in Appendix~\ref{app:meshresults} and Appendix~\ref{app:pcresults} (\Cref{tab:500results,tab:pcresults_geometry}), and standardized bits/input-element rates in Appendix~\ref{app:metrics} (\Cref{tab:standardized_rates}). For all the baselines we compare against, we report average compression ratios, as well as compression and decompression times. The combined forward and reverse mapping networks range from 2.11M to 134.53M parameters (Appendix~\ref{app:training}, \Cref{tab:params}).

\subsection{3D Mesh Compression}
\label{sec:meshcomp}

\begin{figure*}[tb]
    \centering
    \includegraphics[width=\linewidth]{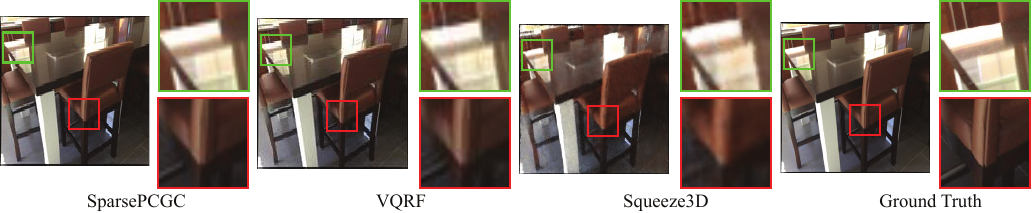}
    \caption{\textbf{Qualitative radiance field compression results.} We show qualitative results comparing \acronym\ to state-of-the-art methods. Our approach achieves a significantly higher compression ratio while maintaining visually important geometric details.}
    \label{fig:rfresults}
\end{figure*}

We compare \acronym\ applied to mesh compression with existing approaches in~\Cref{tab:results} on our test sets. We additionally report strict 3D geometric distortion metrics (Chamfer Distance and EMD) alongside wall-clock times in Appendix~\ref{app:meshresults} (\Cref{tab:500results}), where \acronym\ attains a Chamfer Distance of $0.393$ and an EMD of $0.0084$; these complement the perceptual metrics and confirm that the decompressed geometry matches the input rather than merely looking similar in rendered views. We make the following observations.

First, \acronym\ achieves a \emph{mean} compression ratio of
\textbf{2187\,$\times$} ($6.43\,\mathrm{MB}\!\rightarrow\!3\,\mathrm{kB}$),
more than an order of magnitude higher than that of
DeepSDF~\citep{park2019deepsdf}
(131\,$\times$, $6.43\,\mathrm{MB}\!\rightarrow\!49\,\mathrm{kB}$).
Despite this extreme compactness, \acronym\ preserves perceptual quality,
achieving an LPIPS of $0.0274$ versus $0.3704$ for DeepSDF which completely fails to reconstruct complex large meshes.

\begin{wrapfigure}{r}{0.6\textwidth}
    \vspace{\baselineskip}
    \centering
    \includegraphics[width=\linewidth]{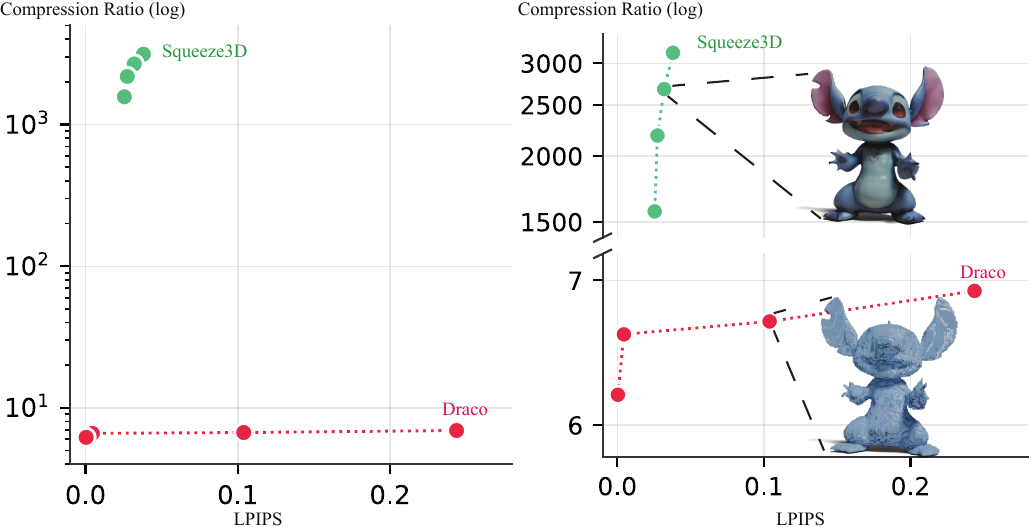}
    \caption{\pink{\textbf{Compression Ratio and Reconstruction Quality Trade-off.} We show the trade-off between compression ratio and reconstruction quality (LPIPS) for \acronym\ and Draco. Each \acronym\ operating point varies $d_C$ and uses a separately trained forward/reverse mapping pair. \acronym\ maintains high quality even at extreme compression ratios, whereas Draco's quality degrades significantly as its compression ratio increases.}}
    \label{fig:pareto}
    \vspace{-\baselineskip}
\end{wrapfigure}
Second, \acronym\ achieves better reconstruction quality (LPIPS of 0.0274) than approaches such as Draco$^\ast$ (LPIPS of 0.2437), Draco$^\dagger$ (LPIPS of 0.1039), and Corto~\citep{corto} (LPIPS of 0.1374). \acronym\ also achieves better quality than Neural Subdivision~\citep{liu2020neural} and DeepSDF~\citep{park2019deepsdf}. Though compared to the state-of-the-art non-learned mesh compression, \acronym\ cannot achieve as high reconstruction quality, we note that \acronym\ achieves a significantly higher compression ratio than these approaches. Neural Geometry Fields~\citep{10.1145/3641519.3657399} performs better in terms of quality and compression ratios than non-learned methods but does significantly worse in terms of compression size when compared with our approach. We conclude that in achieving very high compression rates, \acronym\ offers the highest reconstruction quality. Thus, we demonstrate that leveraging 3D generative models is a promising approach for 3D compression.

While using \acronym\ may not be as fast as some non-learned approaches like Draco~\citep{galligan2018google} or Corto~\citep{corto}, \acronym\ is often faster than other learned methods. \acronym\ is particularly much faster than training a network per object, like in NGF~\citep{10.1145/3641519.3657399}. NGF takes on average 152638 ms to compress objects and 507 ms to decompress objects, compared to 270 ms to compress an object and 1476 ms to decompress objects for \acronym. We qualitatively compare against Draco, Corto, and NGF for compressing meshes in~\Cref{fig:results}. We note that the results from our approach retain high visual quality due to the use of priors from a generative model.

\begin{table*}[tb]
    \caption{\textbf{Mesh Compression.} Quantitative comparison of \acronym\ with state-of-the-art 3D mesh compression methods. We report compression ratio (CR) and quality metrics (PSNR, MS-SSIM, and LPIPS). $\mathcal{T}$ denotes our Objaverse test set, and $\mathcal{O}$ denotes the separately collected set.}
    \centering
    \begin{adjustbox}{width=\textwidth}
    \begin{tabular}{lrrrrrrr}
        \toprule
        \rowcolor{blue!15}Method & CR ($\times$) {\textbf{\scriptsize\textcolor{gray}{(MB)}}} $\uparrow$ & $\mathcal{T}$-PSNR $\uparrow$ & $\mathcal{T}$-MS-SSIM $\uparrow$ & $\mathcal{T}$-LPIPS $\downarrow$ & $\mathcal{O}$-PSNR $\uparrow$ & $\mathcal{O}$-MS-SSIM $\uparrow$ & $\mathcal{O}$-LPIPS $\downarrow$\\
        \midrule
        Draco$^\ast$~\citep{galligan2018google} & 6.92 {\textbf{\scriptsize\textcolor{gray}{(6.43 / 0.93)}}} & 15.45 {\textbf{\scriptsize\textcolor{gray}{($\pm$1.27)}}} & 0.7468 {\textbf{\scriptsize\textcolor{gray}{($\pm$0.09)}}} & 0.2437 {\textbf{\scriptsize\textcolor{gray}{($\pm$0.09)}}} & 15.40 {\textbf{\scriptsize\textcolor{gray}{($\pm$1.30)}}} & 0.7450 {\textbf{\scriptsize\textcolor{gray}{($\pm$0.10)}}} & 0.2450 {\textbf{\scriptsize\textcolor{gray}{($\pm$0.10)}}}\\
        Draco$^\dagger$~\citep{galligan2018google} & 6.70 {\textbf{\scriptsize\textcolor{gray}{(6.43 / 0.96)}}} & 23.33 {\textbf{\scriptsize\textcolor{gray}{($\pm$1.32)}}} & 0.9576 {\textbf{\scriptsize\textcolor{gray}{($\pm$0.03)}}} & 0.1039 {\textbf{\scriptsize\textcolor{gray}{($\pm$0.06)}}} & 23.30 {\textbf{\scriptsize\textcolor{gray}{($\pm$1.35)}}} & 0.9570 {\textbf{\scriptsize\textcolor{gray}{($\pm$0.03)}}} & 0.1045 {\textbf{\scriptsize\textcolor{gray}{($\pm$0.06)}}}\\
        Draco$^\ddagger$~\citep{galligan2018google} & 6.61 {\textbf{\scriptsize\textcolor{gray}{(6.43 / 0.97)}}} & \secondbest{38.91} {\textbf{\scriptsize\textcolor{gray}{($\pm$1.30)}}} & \secondbest{0.9992} {\textbf{\scriptsize\textcolor{gray}{($\pm$0.00)}}} & \secondbest{0.0045} {\textbf{\scriptsize\textcolor{gray}{($\pm$0.00)}}} & \secondbest{38.89} {\textbf{\scriptsize\textcolor{gray}{($\pm$1.30)}}} & \secondbest{0.9991} {\textbf{\scriptsize\textcolor{gray}{($\pm$0.00)}}} & \secondbest{0.0046} {\textbf{\scriptsize\textcolor{gray}{($\pm$0.00)}}}\\
        Draco$^\mathsection$~\citep{galligan2018google} & 6.20 {\textbf{\scriptsize\textcolor{gray}{(6.43 / 1.04)}}} & \best{48.55} {\textbf{\scriptsize\textcolor{gray}{($\pm$1.54)}}} & \best{1.0000} {\textbf{\scriptsize\textcolor{gray}{($\pm$0.00)}}} & \best{0.0004} {\textbf{\scriptsize\textcolor{gray}{($\pm$0.00)}}} & \best{48.50} {\textbf{\scriptsize\textcolor{gray}{($\pm$1.55)}}} & \best{1.0000} {\textbf{\scriptsize\textcolor{gray}{($\pm$0.00)}}} & \best{0.0005} {\textbf{\scriptsize\textcolor{gray}{($\pm$0.00)}}}\\
        Corto~\citep{corto} & 45.93 {\textbf{\scriptsize\textcolor{gray}{(6.43 / 0.14)}}} & 20.92 {\textbf{\scriptsize\textcolor{gray}{($\pm$ 2.79)}}} & 0.8619 {\textbf{\scriptsize\textcolor{gray}{($\pm$ 0.08)}}} & 0.1374 {\textbf{\scriptsize\textcolor{gray}{($\pm$ 0.08)}}} & 20.85 {\textbf{\scriptsize\textcolor{gray}{($\pm$2.80)}}} & 0.8610 {\textbf{\scriptsize\textcolor{gray}{($\pm$0.08)}}} & 0.1380 {\textbf{\scriptsize\textcolor{gray}{($\pm$0.08)}}}\\
        Neural Subd.~\citep{liu2020neural} & 11.28 {\textbf{\scriptsize\textcolor{gray}{(6.43 / 0.57)}}} & 15.95 {\textbf{\scriptsize\textcolor{gray}{($\pm$2.18)}}} & 0.8525 {\textbf{\scriptsize\textcolor{gray}{($\pm$0.04)}}} & 0.1513 {\textbf{\scriptsize\textcolor{gray}{($\pm$0.05)}}} & 15.90 {\textbf{\scriptsize\textcolor{gray}{($\pm$2.20)}}} & 0.8520 {\textbf{\scriptsize\textcolor{gray}{($\pm$0.04)}}} & 0.1520 {\textbf{\scriptsize\textcolor{gray}{($\pm$0.05)}}}\\
        DeepSDF~\citep{park2019deepsdf} & 131.22 {\textbf{\scriptsize\textcolor{gray}{(6.43 / 0.05)}}} & 8.47 {\textbf{\scriptsize\textcolor{gray}{($\pm$ 0.23)}}} & 0.7039 {\textbf{\scriptsize\textcolor{gray}{($\pm$ 0.07)}}} & 0.3704 {\textbf{\scriptsize\textcolor{gray}{($\pm$ 0.08)}}} & 8.21 {\textbf{\scriptsize\textcolor{gray}{($\pm$0.25)}}} & 0.6850 {\textbf{\scriptsize\textcolor{gray}{($\pm$0.08)}}} & 0.3912 {\textbf{\scriptsize\textcolor{gray}{($\pm$0.09)}}}\\
        NGF~\citep{10.1145/3641519.3657399} & 42.87 {\textbf{\scriptsize\textcolor{gray}{(6.43 / 0.15)}}} & 35.45 {\textbf{\scriptsize\textcolor{gray}{($\pm$ 3.02)}}} & 0.9987 {\textbf{\scriptsize\textcolor{gray}{($\pm$ 0.08)}}} & 0.0054 {\textbf{\scriptsize\textcolor{gray}{($\pm$ 0.03)}}} & 35.40 {\textbf{\scriptsize\textcolor{gray}{($\pm$3.05)}}} & 0.9985 {\textbf{\scriptsize\textcolor{gray}{($\pm$0.08)}}} & 0.0056 {\textbf{\scriptsize\textcolor{gray}{($\pm$0.03)}}}\\
        3DShape2VecSet~\citep{zhang20233dshape2vecset} & \secondbest{342.50} {\textbf{\scriptsize\textcolor{gray}{(6.43 / 0.0188)}}} & 21.65 {\textbf{\scriptsize\textcolor{gray}{($\pm$2.22)}}} & 0.9047 {\textbf{\scriptsize\textcolor{gray}{($\pm$0.06)}}} & 0.1582 {\textbf{\scriptsize\textcolor{gray}{($\pm$0.09)}}} & 21.91 {\textbf{\scriptsize\textcolor{gray}{($\pm$2.55)}}} & 0.9220 {\textbf{\scriptsize\textcolor{gray}{($\pm$0.04)}}} & 0.1547 {\textbf{\scriptsize\textcolor{gray}{($\pm$0.07)}}}\\
        \midrule
        \acronym\ (InstantMesh) & \best{2187.07} {\textbf{\scriptsize\textcolor{gray}{(6.43 / 0.003)}}} & 27.50 {\textbf{\scriptsize\textcolor{gray}{($\pm$3.13)}}} & 0.9796 {\textbf{\scriptsize\textcolor{gray}{($\pm$0.02)}}} & 0.0274 {\textbf{\scriptsize\textcolor{gray}{($\pm$0.02)}}} & 26.95 {\textbf{\scriptsize\textcolor{gray}{($\pm$3.20)}}} & 0.9905 {\textbf{\scriptsize\textcolor{gray}{($\pm$0.01)}}} & 0.0124 {\textbf{\scriptsize\textcolor{gray}{($\pm$0.01)}}}\\
        \bottomrule
    \end{tabular}
    \end{adjustbox}
    \label{tab:results}
\end{table*}

\subsection{Compression Ratio and Reconstruction Quality Trade-off}
\label{sec:pareto}

To better understand the trade-off between compression ratio and reconstruction quality, we analyze the performance of \acronym\ against Draco~\citep{galligan2018google} across different compression settings. While many neural compression methods do not offer tunable compression ratios, \acronym\ allows for different compression levels by varying the size of the compressed latent representation. Similarly, Draco offers tunable compression through its quantization parameters. In~\Cref{fig:pareto}, we plot the Pareto frontier for both methods, showing the relationship between compression ratio (x-axis) and reconstruction quality (y-axis, measured by LPIPS). We observe that Draco~\citep{galligan2018google} can only achieve compression ratios which are orders of magnitude lower than \acronym, and as its compression ratio increases, its quality degrades significantly. In contrast, \acronym\ maintains high visual quality. We further analyze the effect of the compressed-latent size on quality in Appendix~\ref{app:ablations} (\Cref{tab:ablations}).

\pink{For applications where fine geometric details are critical, such as CAD and cultural heritage, \acronym\ may be less suitable; for applications prioritizing storage or bandwidth, such as streaming and asset libraries, it offers a compelling option.}
\pink{We also vary $d_C$ for point clouds and report bits per input point versus PointSSIM in Appendix~\ref{app:pcresults} (\Cref{fig:pc_pareto_sparsepcgc}). Each point on the \acronym\ curves uses its own trained mapping pair. We do not report an analogous radiance-field sweep: increasing $d_C$ substantially enlarges the 3D convolutional mapping networks, and training enough larger operating points was computationally infeasible in our setup. Point-cloud mapping networks are smaller, making this sweep practical.}
Beyond storage, mapping into the generator's \emph{semantic} latent space yields a representation that supports operations unavailable to standard codecs such as Draco: as shown in Appendix~\ref{app:meshresults} (\Cref{fig:interp}), the compressed codes of two objects can be linearly interpolated to produce smooth, plausible intermediate shapes, suggesting the learnt compressed space is well-structured and could enable downstream tasks such as 3D asset mixing.

\subsection{3D Point Cloud Compression}

We compare our method applied to 3D point cloud compression with previous approaches in~\Cref{tab:pcresults}.~\Cref{tab:pcresults_geometry} additionally reports strict 3D geometric distortion metrics (Chamfer Distance of $0.0285$ and EMD of $0.0062$). We notice that our approach achieves a significantly higher compression ratio of 58.50$\times$ (117\,KB / 2.00\,KB) but leads to a lower PCQM.

\begin{wrapfigure}[13]{R}{0.45\textwidth}
    \centering
    \includegraphics[width=\linewidth]{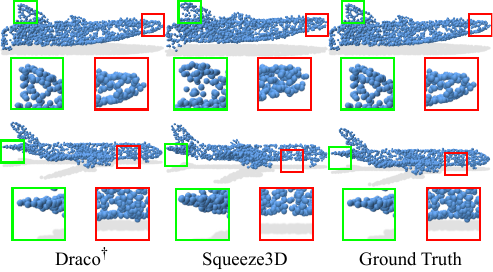}
    \caption{\textbf{Qualitative point-cloud compression.} Reconstructions from \acronym\ and established codecs; quantitative results are reported in~\Cref{tab:pcresults,fig:pc_pareto_sparsepcgc}.}
    \label{fig:otheresults}
\end{wrapfigure}

We demonstrate that our framework can be used off-the-shelf for point clouds but might need further improvements to match the quality of other methods. We show qualitative results in~\Cref{fig:otheresults}. We note that the Draco results shown in~\Cref{fig:otheresults} correspond to a much higher-storage operating point; \acronym's advantage lies in the extreme-compression regime, where Draco's quality degrades sharply as its ratio increases (\Cref{fig:pc_pareto_sparsepcgc}).

\subsection{Radiance Field Compression}
\label{sec:radiancecomp}

We compare \acronym\ applied to radiance field compression with SparsePCGC~\citep{9968173} and VQRF~\citep{10204677} in~\Cref{tab:rfresults}. We choose these approaches to compare against since these works (or a setting of these works), akin to our setup, only require (rgb$\sigma$) grids. 
We notice that our approach achieves a significantly higher compression ratio of 619$\times$ as opposed to 40$\times$ for VQRF. \acronym\ achieves these significantly higher compression rates with only a 0.0125 increase in LPIPS. We show qualitative results in~\Cref{fig:rfresults}.

\begin{table}[b]
    \centering
    \caption{\textbf{Point Cloud Compression.} Quantitative comparison of \acronym\ with state-of-the-art point cloud compression methods. We report compression ratio (CR), compression and decompression times, and quality metrics (PCQM and PointSSIM).
    }
    \begin{adjustbox}{width=\textwidth}
    \begin{tabular}{lrrrrr}
        \toprule\rowcolor{blue!15}Method & CR ($\times$) {\textbf{\scriptsize\textcolor{gray}{(KB)}}} $\uparrow$ & Compress (ms) $\downarrow$ & Decompress (ms) $\downarrow$ & PCQM $\uparrow$ & PointSSIM $\uparrow$\\
        \midrule Draco$^\ddagger$~\citep{galligan2018google} & 15.64 {\textbf{\scriptsize\textcolor{gray}{(117 / 7.48)}}} & \secondbest{0.35} & \secondbest{0.34} & \secondbest{3.2875} {\textbf{\scriptsize\textcolor{gray}{($\pm$ 0.13)}}} & \best{0.9722} {\textbf{\scriptsize\textcolor{gray}{($\pm$ 0.11)}}}\\
        Draco$^\mathsection$~\citep{galligan2018google} & 22.41 {\textbf{\scriptsize\textcolor{gray}{(117 / 5.22)}}} & \best{0.25} & \best{0.23} & 2.1039 {\textbf{\scriptsize\textcolor{gray}{($\pm$ 0.08)}}} & \secondbest{0.9535} {\textbf{\scriptsize\textcolor{gray}{($\pm$ 0.12)}}}\\
        G-PCC~\citep{schwarz2018emerging} & 37.21 {\textbf{\scriptsize\textcolor{gray}{(117 / 3.15)}}} & 19.76 & 29.92 & 2.8163 {\textbf{\scriptsize\textcolor{gray}{($\pm$ 0.10)}}} &  0.9331 {\textbf{\scriptsize\textcolor{gray}{($\pm$ 0.09)}}}\\
        V-PCC~\citep{liu2019comprehensive} & \secondbest{50.17} {\textbf{\scriptsize\textcolor{gray}{(117 / 2.33)}}} & 2.33 & 234.85 & \best{48.22} {\textbf{\scriptsize\textcolor{gray}{($\pm$0.13)}}} & 0.9437 {\textbf{\scriptsize\textcolor{gray}{($\pm$0.09)}}}\\
        SparsePCGC~\citep{9968173} & 34.37 {\textbf{\scriptsize\textcolor{gray}{(117 / 3.55)}}} & 121.35 & 131.82 & 0.0653 & 0.1909\\
        \midrule \acronym\ (512, LION) & \best{58.50} {\textbf{\scriptsize\textcolor{gray}{(117 / 2.00)}}} & 3.85 & 12.74 & 1.8437 {\textbf{\scriptsize\textcolor{gray}{($\pm$1.13)}}} & 0.4484 {\textbf{\scriptsize\textcolor{gray}{($\pm$0.11)}}}\\
        \bottomrule
    \end{tabular}
    \end{adjustbox}
    \label{tab:pcresults}
\end{table}
\begin{table}[t]
    \centering
    \caption{\textbf{Radiance Field Compression.} Quantitative comparison of \acronym\ with state-of-the-art 3D radiance field compression methods. We report compression ratio (CR), compression and decompression times, and quality metrics (PSNR, MS-SSIM, and LPIPS).
    }
    \begin{adjustbox}{width=\textwidth}
    \begin{tabular}{lrrrrrr}
        \toprule
        \rowcolor{blue!15}Method & CR ($\times$) {\textbf{\scriptsize\textcolor{gray}{(MB)}}} $\uparrow$ & Compress (ms) $\downarrow$ & Decompress (ms) $\downarrow$ & PSNR $\uparrow$ & MS-SSIM $\uparrow$ & LPIPS $\downarrow$\\
        \midrule
        SparsePCGC~\citep{9968173} & \secondbest{78.92} {\textbf{\scriptsize\textcolor{gray}{(58.07 / 0.74)}}} & 301.39 & 680.82 & 22.26 {\textbf{\scriptsize\textcolor{gray}{$\pm$ 0.92}}} & 0.8947 {\textbf{\scriptsize\textcolor{gray}{$\pm$ 0.02}}} & 0.1400 {\textbf{\scriptsize\textcolor{gray}{$\pm$ 0.02}}}\\
        VQRF~\citep{10204677} & 40.25 {\textbf{\scriptsize\textcolor{gray}{(58.07 / 1.45)}}} & \secondbest{120.14} & \best{20.58} & \best{29.55} {\textbf{\scriptsize\textcolor{gray}{$\pm$ 0.01}}} & \best{0.9749} {\textbf{\scriptsize\textcolor{gray}{$\pm$ 0.00}}} & \best{0.0618} {\textbf{\scriptsize\textcolor{gray}{$\pm$ 0.00}}} \\
        \midrule
        \acronym\ (NeRF-MAE) & \best{619.41} {\textbf{\scriptsize\textcolor{gray}{(58.07 / 0.09)}}} & \best{45.63} & \secondbest{75.80} & \secondbest{26.62} {\textbf{\scriptsize\textcolor{gray}{$\pm$ 2.57}}} & \secondbest{0.9533} {\textbf{\scriptsize\textcolor{gray}{$\pm$ 0.04}}} & \secondbest{0.0743} {\textbf{\scriptsize\textcolor{gray}{$\pm$ 0.02}}}\\
        \bottomrule
    \end{tabular}
    \end{adjustbox}
    \label{tab:rfresults}
\end{table}

\subsection{Comparison with Native Generative Model Encoders}
\label{sec:native_baseline}

We perform an additional experiment to compare \acronym\ against a baseline that uses the native encoder and decoder of the generative models themselves for compression. This baseline effectively uses the latent space of the generative model as the compressed representation, and therefore serves as the closest available proxy for the quality of the ``uncompressed'' generator latent: it quantifies how much quality \acronym\ gives up to obtain its much smaller code. (Note there is no single uncompressed latent to compare against directly, since the encoder and generator operate in different, differently sized latent spaces; see Appendix~\ref{app:training}, \Cref{tab:latentsizes}.) For reference, in the Shap-E setting (\Cref{tab:native_shape}) the input mesh is encoded by MeshAnything into a $257\times1024$ latent, which \acronym\ compresses to a $1024$-dimensional code before mapping it into the Shap-E decoder latent (selected encoder and generator latent sizes are listed in Appendix~\ref{app:training}, \Cref{tab:latentsizes}). Since these generative models are designed for high-fidelity generation rather than compression, their latent spaces are typically high-dimensional.

We perform this comparison for Shap-E~\citep{jun2023shapegeneratingconditional3d}, LION~\citep{zeng2022lion}, and NeRF-MAE~\citep{irshad2024nerf}. We do not include InstantMesh~\citep{xu2024instantmesh} or LRM~\citep{hong2024lrmlargereconstructionmodel} in this comparison because these are feed-forward image-to-3D models that utilize 2D image encoders (e.g., DINO). They lack a native 3D-to-3D encoder that can map a 3D mesh directly to their latent space, making them unsuitable for this specific 3D compression baseline without introducing a lossy intermediate rendering step.

We report the results in~\Cref{tab:native_shape,tab:native_lion,tab:native_nerfmae}. We observe that the native baselines generally achieve high reconstruction quality (serving as an upper bound) but at significantly lower compression ratios. In contrast, \acronym\ achieves orders of magnitude higher compression ratios while maintaining comparable visual quality. In some cases, such as with Shap-E, our method can even slightly outperform the native baseline on certain metrics.

\begin{table}[tb]
    \centering
    \caption{\pink{\textbf{Shap-E Comparison.} Comparison between Squeeze3D (MeshAnything $\to$ Shap-E) and the native Shap-E encoder-decoder.}}
    \begin{tabular}{lrrrr}
        \toprule
        \rowcolor{blue!15}Method & CR ($\times$) $\uparrow$ & PSNR $\uparrow$ & MS-SSIM $\uparrow$ & LPIPS $\downarrow$ \\
        \midrule
        Shap-E Native Baseline & \secondbest{1.61} & \secondbest{25.80} & \secondbest{0.9600} & \secondbest{0.0410} \\
        \acronym\ (Shap-E) & \best{1607.50} & \best{25.92} & \best{0.9615} & \best{0.0398} \\
        \bottomrule
    \end{tabular}
    \label{tab:native_shape}
\end{table}

\begin{table}[t]
    \centering
    \caption{\textbf{LION Comparison.} Comparison between \acronym\ (PointNet++ $\to$ LION) and the native LION encoder-decoder.}
    \begin{tabular}{lrrrr}
        \toprule
        \rowcolor{blue!15}Method & CR ($\times$) $\uparrow$ & PCQM $\uparrow$ & PointSSIM $\uparrow$ & Chamfer $\downarrow$ \\
        \midrule
        LION Native Baseline & \secondbest{3.60} & \best{2.1000} & \best{0.5100} & \best{0.0210} \\
        \acronym\ (LION) & \best{58.50} & \secondbest{1.8437} & \secondbest{0.4484} & \secondbest{0.0285} \\
        \bottomrule
    \end{tabular}
    \label{tab:native_lion}

    \vspace{\floatsep}
    \caption{\textbf{NeRF-MAE Comparison.} Comparison between \acronym\ (NeRF-MAE $\to$ NeRF-MAE) and the native NeRF-MAE encoder-decoder.}
    \begin{tabular}{lrrrr}
        \toprule
        \rowcolor{blue!15}Method & CR ($\times$) $\uparrow$ & PSNR $\uparrow$ & MS-SSIM $\uparrow$ & LPIPS $\downarrow$ \\
        \midrule
        NeRF-MAE Native Baseline & \secondbest{7.44} & \best{28.10} & \best{0.9700} & \best{0.0550} \\
        \acronym\ (NeRF-MAE) & \best{619.41} & \secondbest{26.62} & \secondbest{0.9533} & \secondbest{0.0743} \\
        \bottomrule
    \end{tabular}
    \label{tab:native_nerfmae}
\end{table}

\Needspace{8\baselineskip}
\subsection{Decoder Cost and Amortization}
\label{sec:amortization}

Following standard compression conventions, our reported ratios do not include the size of the pre-trained decoder, just as a codec is not counted toward the size of each compressed file. The decoder in \acronym\ is a general, object-agnostic asset, stored once and reused to decompress every object and usable for other tasks such as generation, so we do not count it toward the compressed size. We list the total decoder sizes in~\Cref{tab:amortize}. \pink{These one-time costs can be amortized after hundreds of objects for meshes, tens for radiance fields, and thousands for point clouds.}

\section{Discussion and Limitations}
\label{sec:discussion}

\pink{The most significant limitation of our approach is its inherent dependency on the quality and expressiveness of the underlying 3D generative model. The decompressed outputs from \acronym\ fundamentally cannot exceed the quality of what the generator can produce, as the generator serves as both a prior knowledge base and a quality ceiling. In our evaluation, we observed that reconstruction fidelity is directly correlated with the generative capabilities of the chosen generator model.}

\begin{wraptable}[14]{r}{0.4\textwidth}
    \centering
    \vspace{-2em}
    \caption{Storage size of each decoder used at decompression.}
    \begin{tabular}{lr}
        \toprule
        \rowcolor{blue!15}Decoder & Total size \\
        \midrule
        \rowcolor{gray!15}\multicolumn{2}{l}{\acronym} \\
        Shap-E      & 905\,MB  \\
        InstantMesh & 1.51\,GB \\
        OpenLRM     & 1.04\,GB \\
        LION        & 270\,MB  \\
        NeRF-MAE    & 306\,MB  \\
        \midrule
        \rowcolor{gray!15}\multicolumn{2}{l}{Baselines} \\
        3DShape2VecSet & 389\,MB \\
        DeepSDF        & 7.4\,MB \\
        \bottomrule
    \end{tabular}
    \label{tab:amortize}
\end{wraptable}
\noindent\pink{For instance, when using the Shap-E~\citep{jun2023shapegeneratingconditional3d} generator, we found it challenging to faithfully reproduce highly complex objects due to limitations in the model's semantic capacity (see \Cref{sec:failure_cases}). Concretely, geometric detail that lies beyond the generator's representational capacity, such as fine threads on a screw or small embossed text, is smoothed away or hallucinated rather than reconstructed faithfully (\Cref{sec:failure_cases}); the generator acts as a hard quality ceiling. This limitation, however, positions \acronym\ to benefit automatically from future advancements in 3D generative modeling. Newer generators such as Hunyuan3D-2.5~\citep{lai2025hunyuan3d25highfidelity3d} and TRELLIS~\citep{xiang2024structured,xiang2025trellis2} may provide broader priors and higher-fidelity reconstruction. Their larger latent spaces may require wider mapping networks and additional compute.}

\newpage
\pink{Since \acronym\ is designed as a compression-decompression framework that should not require per-scene retraining, the generalization capability of our mapping networks to unseen 3D models is crucial. We evaluated this aspect by testing on a dataset of 500 3D meshes and 227 radiance fields, which served as out-of-distribution samples, and found that our approach maintains consistent performance (see \Cref{app:results}). Nevertheless, for certain outlier cases where the input 3D model contains features far from the distribution seen during training, compression quality may degrade. Encoder-limited failures may benefit from multiscale encoders~\citep{dagli2026adavomp} or higher-resolution input sampling, while generator-limited failures require a broader prior or a conventional fallback codec. A practical deployment could use the native encoder's reconstruction error or an OOD detector to identify risky inputs and route them to a standard codec.}

\section{Conclusion}
\label{sec:conclusion}

In this work, we answer: \textit{How can we bridge the latent spaces of different models?} We do so with losses in the latent space, without any reconstruction-based losses. As an application, we apply this to 3D. We introduce \acronym, a framework for 3D compression that leverages the rich priors in existing 3D generation models, bridging arbitrary latent spaces between different models to enable extreme compression ratios while maintaining high visual fidelity. By mapping between latent spaces, we avoid the per-scene optimization that plagues many neural compression methods, yielding faster compression and decompression than learned approaches that train a network per object (\eg, NGF), at competitive visual quality, though the wall-clock time remains dominated by the chosen encoder and decoder. We hope this work inspires more research on using generative models for compressing 3D data.

\section*{Broader Impact Statement}

\pink{Compression with generative priors can reconstruct geometry that is perceptually plausible but not faithful to the original: fine details, text, and structures outside the generator's prior may be smoothed, altered, or hallucinated. Consequently, \acronym\ should not be used where exact geometry is safety-critical without fidelity checks and a fallback to a conventional codec or the original asset.}

\ifdefined\anonymoussubmission
\else
\section*{Acknowledgements}

Resources used in preparing this research were provided in part by the Province of Ontario, the Government of Canada through CIFAR, and companies sponsoring the Vector Institute.
\fi

\newpage
\nocite{polyscope}

\bibliographystyle{tmlr}
\bibliography{main}

\newpage

\appendix
\makeatletter
\setlength{\@fptop}{0pt}
\setlength{\@fpsep}{\floatsep}
\setlength{\@fpbot}{0pt plus 1fil}
\setlength{\@dblfptop}{0pt}
\setlength{\@dblfpsep}{\floatsep}
\setlength{\@dblfpbot}{0pt plus 1fil}
\makeatother
\setcounter{topnumber}{3}
\section*{Appendix Contents}
\addcontentsline{toc}{section}{Appendix Contents}
\begingroup
\setcounter{tocdepth}{2}
\startcontents[appendix]
\printcontents[appendix]{l}{1}{\setcounter{tocdepth}{2}}
\endgroup

\newpage

\section{Additional Implementation Details}
\label{app:additional}

We share additional implementation details to reproduce \acronym.

\subsection{Network Design}
\label{app:networkdesign}

\begin{wrapfigure}{r}{0.5\textwidth} \centering
    \vspace{-4em} \includegraphics[width=\linewidth]{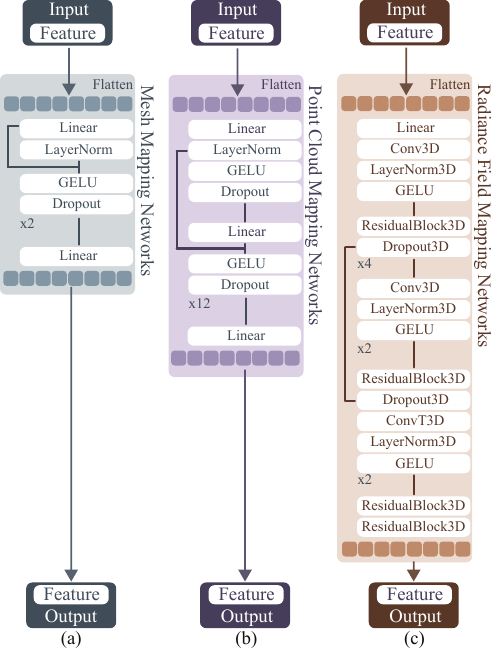}
    \caption{\textbf{Network Architectures.}}
    \label{fig:arch}
    \vspace{-4em} \end{wrapfigure}

For each of the encoder-decoder pairs, we train the mapping networks, which are feed-forward neural networks. We summarize network architectures in~\Cref{fig:arch}. Our architectures for different representations are highly similar.

\paragraph{Meshes.} The mapping networks together first flatten the input and apply a linear layer to project it into a hidden dimension. This is followed by LayerNorm~\citep{ba2016layernormalization}, a GELU nonlinearity~\citep{hendrycks2023gaussianerrorlinearunits}, and dropout~\citep{JMLR:v15:srivastava14a}. A second linear layer then produces another hidden representation, which is added residually to the output of the first linear layer and passed through a second LayerNorm. After another GELU and dropout, a final linear transformation projects into the latent space.

\paragraph{Point Clouds.} The mapping networks together consist of flattening the input and feeding it through a linear projection and LayerNorm~\citep{ba2016layernormalization}; this normalised vector is stored as a global residual. The sequence then passes through GELU-activated~\citep{hendrycks2023gaussianerrorlinearunits} hidden layers of uniform width, each followed by dropout regularisation on the activations~\citep{JMLR:v15:srivastava14a}. We use two skip-connection schemes: (i) the global residual is re-added immediately after the first hidden layer, and (ii) every fourth hidden layer receives a local residual that adds its own input to its output, creating short four-layer paths. A final linear projection transforms the resulting representation into the target latent space.

\paragraph{Radiance Fields.} The mapping networks together consist of a custom LayerNorm that first permutes tensors so that LayerNorm~\citep{ba2016layernormalization} can operate across channels before restoring the usual NCDHW layout. We then build ResidualBlocks, which mirror a bottleneck ResNet-V2 design~\citep{10.1007/978-3-319-46493-0_38}: a $3\times3\times3$ conv–LN–GELU \citep{hendrycks2023gaussianerrorlinearunits} stem, a $1\times1\times1 \rightarrow 3\times3\times3$ bottleneck branch that doubles the channel count, and two skip paths: (i) projection shortcut for stride/width mismatches and (ii) “within-block” residual that re-adds the pre-bottleneck activation just before the final GELU. The encoder begins with a 96-channel input, expands to 192 channels, and applies two strided ResidualBlock units to down-sample from $40^3$ latents to $20^3$ and then $10^3$. Thus we have a narrow bottleneck of 24 channels that serves as the latent code. The symmetric decoder inverts this pathway with a $3\times3\times3$ convolution, residual processing, and two transposed-conv up-sampling stages that restore the original resolution, all regularised by spatial Dropout in 3D. U-Net–style~\citep{10.1007/978-3-319-24574-4_28} skip connections add encoder feature maps to decoder activations whenever spatial dimensions match.

In all cases the reverse mapping network produces a single flat vector whose dimensionality equals the flattened size of the target decoder's latent, and a fixed reshape maps it to the structure the decoder expects. Thus, we use no decoder-specific architectures. For InstantMesh and LRM, the decoder takes a triplane, so the final linear layer produces $3\times80\times64\times64$ values that are reshaped into the $(3, 80, 64, 64)$ triplane tensor before being passed to the frozen triplane decoder. For LION, whose latent is an unordered set rather than a grid, the latent is a fixed-size concatenation of a global shape code and a per-point local code; the network outputs an $8320$-dimensional vector that we split into the first $128$ dimensions (the global latent, reshaped to $1\times128\times1\times1$) and the remaining $8192$ dimensions (the local latent, reshaped to $1\times8192\times1\times1$). Because the local code has a fixed length and canonical ordering, the mapping network never has to predict a variable-size or permutation-invariant point set directly.

\FloatBarrier
\subsection{Training}
\label{app:training}

\begin{table}[tb]
    \centering
    \caption{\textbf{Model Sizes.} Combined parameter count of the forward and reverse mapping networks trained for each setting.}
    \begin{adjustbox}{width=\textwidth}
    \begin{tabular}{lr}
        \toprule \rowcolor{blue!15}Encoder-Decoder Pair & Parameters (M)\\
        \midrule MeshAnything~\citep{chen2024meshanythingartistcreatedmeshgeneration}-InstantMesh~\citep{xu2024instantmesh} & 96.12 \\
        MeshAnything~\citep{chen2024meshanythingartistcreatedmeshgeneration}-Open LRM~\citep{hong2024lrmlargereconstructionmodel, openlrm} & 87.51 \\
        MeshAnything~\citep{chen2024meshanythingartistcreatedmeshgeneration}-Shap-E~\citep{jun2023shapegeneratingconditional3d} & 134.53 \\
        \midrule PointNet++~\citep{NIPS2017_d8bf84be}-LION~\citep{zeng2022lion} (512/1024) & 2.11 \\
        PointNet++~\citep{NIPS2017_d8bf84be}-LION~\citep{zeng2022lion} (2048) & 6.53\\
        PointNet++~\citep{NIPS2017_d8bf84be}-LION~\citep{zeng2022lion} (4096) & 22.29\\
        PointNet++~\citep{NIPS2017_d8bf84be}-LION~\citep{zeng2022lion} (8192) & 81.48\\
        \midrule NeRF-MAE~\citep{irshad2024nerf} & 86.46\\
        \bottomrule
    \end{tabular}
    \end{adjustbox}
    \label{tab:params}
\end{table}

\begin{table*}[tb]
    \centering
    \caption{\textbf{Latent Space Sizes.} Sizes of latent spaces for encoder and generator models we use.}
    \begin{tabular}{lr}
    \toprule \rowcolor{blue!15}Encoder/Generator Model & Latent Space Size\\
    \midrule MeshAnything~\citep{chen2024meshanythingartistcreatedmeshgeneration} & $257\times 1024$\\
    InstantMesh TriPlane Features~\citep{xu2024instantmesh} & $3\times80\times64\times64$\\
    LION~\citep{zeng2022lion} &	8320\\
    PointNet++~\citep{NIPS2017_d8bf84be} &	1024\\
    NeRF-MAE~\citep{irshad2024nerf} & $256\times20\times20\times20$ \\
    \bottomrule
    \end{tabular}
    \label{tab:latentsizes}
\end{table*}

\begin{table}[tb]
    \centering
    \pinkall
    \small
    \caption{\textbf{Latent Sizes.} The sizes of the compressed representations for our experiments.}
    \begin{tabular}{lrr}
        \toprule
        \rowcolor{blue!15}Encoder $\rightarrow$ decoder & $d_C$ & Latent size (KB)\\
        \midrule
        MeshAnything $\rightarrow$ OpenLRM & 1024 & 4.00\\
        MeshAnything $\rightarrow$ Shap-E & 1024 & 4.00\\
        MeshAnything $\rightarrow$ InstantMesh & 770 & 3.01\\
        PointNet++ $\rightarrow$ LION & 512 & 2.00\\
        PointNet++ $\rightarrow$ LION & 1024 & 4.00\\
        PointNet++ $\rightarrow$ LION & 2048 & 8.00\\
        PointNet++ $\rightarrow$ LION & 4096 & 16.00\\
        PointNet++ $\rightarrow$ LION & 8192 & 32.00\\
        NeRF-MAE $\rightarrow$ NeRF-MAE & 24000 & 93.75\\
        \bottomrule
    \end{tabular}
    \label{tab:compressed_latent_sizes}
\end{table}

Our results are collected on a machine with either $1\times$NVIDIA RTX 4090 GPU or $1-4\times$ H100 GPUs. We list the choices we make during training for each encoder-generator pair.

\begin{table}[t]
    \centering
    \pinkall
    \caption{\textbf{One-time Training and Data-creation Costs.} These costs are incurred once for the listed encoder-generator setting; subsequent objects use feed-forward inference.}
    \begin{tabular}{lrr}
        \toprule
        \rowcolor{blue!15}Decoder & Mapping training (h) & Paired-data creation (h)\\
        \midrule
        InstantMesh & 12 & 20\\
        LION & 5 & 20\\
        NeRF-MAE & 15 & 3\\
        \bottomrule
    \end{tabular}
    \label{tab:training_costs}
\end{table}

For all the models we train with MeshAnything~\citep{chen2024meshanythingartistcreatedmeshgeneration} as the encoder, we use the 350-million-parameter version of MeshAnything and use the features created by MeshAnything as the encoded inputs. These encoded inputs \texttt{(257, 1024)} are relatively larger compared to our desired compression size.

\paragraph{MeshAnything~\citep{chen2024meshanythingartistcreatedmeshgeneration}-InstantMesh~\citep{xu2024instantmesh}.} While generating the dataset, we ensure that the input condition images are in the RGBA format and have no background. We experiment with having the mapping networks generate multiview ViT embeddings~\citep{dosovitskiy2021an} that InstantMesh uses, as well as the triplane representation InstantMesh uses. We observed better performance when generating triplane representations; thus, we report results for this setting.

\paragraph{MeshAnything~\citep{chen2024meshanythingartistcreatedmeshgeneration}-Open LRM~\citep{hong2024lrmlargereconstructionmodel, openlrm}.} While generating the dataset, we ensure that the input condition images are in the RGBA format and have no background. Due to the lack of public code for the LRM, we use the OpenLRM implementation. We experiment with both \texttt{openlrm-mix-base-1.1} and \texttt{openlrm-obj-base-1.1}, and we observe that \texttt{openlrm-obj-base-1.1} performs better. We train the mapping networks to generate the triplane latents for LRM.

\paragraph{MeshAnything~\citep{chen2024meshanythingartistcreatedmeshgeneration}-Shap-E~\citep{jun2023shapegeneratingconditional3d}.} We experiment with Shap-E in the text-to-3D mode. The mapping networks generate the implicit MLP representations.

\paragraph{PointNet++~\citep{NIPS2017_d8bf84be}-LION~\citep{zeng2022lion}.} We experiment with the \texttt{SSG} and \texttt{MSG} versions of PointNet++ and observe that \texttt{SSG} without normals works best. We use the PointNet++ checkpoints pre-trained for classification. We consider the outputs from the point set feature learning module as the encoded representations. We use the \texttt{all} categories model for LION. The mapping networks are trained to generate both the global and local latents for LION. The same synthetic LION paired-latent dataset is shared across all compressed-dimensionality variants in~\Cref{tab:compressed_latent_sizes}.

We provide the model sizes in~\Cref{tab:params}, compressed latent sizes in~\Cref{tab:compressed_latent_sizes}, one-time measured costs in~\Cref{tab:training_costs}, hyperparameters in~\Cref{tab:hypmesh}, and latent-space sizes in~\Cref{tab:latentsizes}.

\begin{table*}[tb]
\centering
\caption{\textbf{Training Hyperparameters.} We show the training hyperparameters for the mapping networks we train.}
\begin{tabular}{lrrrr}
\toprule
\rowcolor{blue!15}Hyperparameter & LRM & Shap-E & InstantMesh & NeRF-MAE\\
\midrule
Training Precision & FP-32 & FP-32 & FP-32 & FP-32 \\
$d_C$ & 1024 & 1024 & 770 & 24000\\
Dropout & 0.35 & 0.35 & 0.35 & 0.2 \\
Epochs & 700 & 700 & 700 & 200\\
Batch Size & 16 & 8 & 16 & 4\\
Optimizer & Muon & Adam & Muon & Muon \\
\multirow{3}{*}{Optimizer Parameters} & & $\lambda=10^{-2}$ & &\\
& NS = 6 & $\beta_1 = 0.9$ & NS = 6 & NS = 6\\
& $\beta=0.95$ & $\beta_2 = 0.999$ & $\beta=0.95$ & $\beta=0.95$\\
Initial Learning Rate & $10^{-2}$ & $10^{-4}$ & $10^{-3}$ & $10^{-2}$\\
Final Learning Rate & $10^{-7}$ & $10^{-4}$ & $10^{-7}$ & $10^{-5}$\\
Scheduler & Linear Decay & Constant & Linear Decay & Linear Decay \\
Decay Epochs & 700 & N/A & 600 & 100\\
Gradient Accum. Steps & 1 & 2 & 1 & 1\\
Gradient Clipping & None & None & None & None\\
\bottomrule
\end{tabular}
\begin{tabular}{lrrrr}
\toprule
\rowcolor{blue!15}Hyperparameter & LION (512/1024) & LION (2048) & LION (4096) & LION (8192) \\
\midrule
Training Precision & FP-32 & FP-32 & FP-32 & FP-32 \\
$d_C$ & 512/1024 & 2048 & 4096 & 8192 \\
Dropout & 0.3 & 0.3 & 0.3 & 0.3 \\
Epochs & 400 & 400 & 400 & 400 \\
Batch Size & 16 & 16 & 16 & 16 \\
Optimizer & Muon & Muon & Muon & Muon \\
\multirow{2}{*}{Optimizer Parameters} & $\beta=0.95$ & $\beta=0.95$ & $\beta=0.95$ & $\beta=0.95$ \\
& NS = 6 & NS = 6 & NS = 6 & NS = 6 \\
Initial Learning Rate & $10^{-3}$ & $10^{-3}$ & $10^{-3}$ & $10^{-3}$ \\
Final Learning Rate & $10^{-7}$ & $10^{-7}$ & $10^{-7}$ & $10^{-7}$ \\
Scheduler & Linear Decay & Linear Decay & Linear Decay & Linear Decay \\
Decay Epochs & 100 & 100 & 100 & 100 \\
Gradient Accum. Steps & 1 & 1 & 1 & 1 \\
Gradient Clipping & None & None & None & None \\
\bottomrule
\end{tabular}
\label{tab:hypmesh}
\end{table*}

\section{Evaluation Details}
\label{app:eval}

\acronym\ leverages pre-trained generative priors strictly as a decoding mechanism for compressing arbitrary 3D data, distinct from compressing the native outputs of generative models. Direct native encoder-decoder comparisons are possible for Shap-E, LION, and NeRF-MAE, as reported in~\Cref{sec:native_baseline}. They are not available for feed-forward image-to-3D models such as InstantMesh and LRM, which lack native 3D encoders and would require a lossy intermediate rendering step.

\subsection{Metrics}
\label{app:metrics}

For all mesh and radiance-field experiments, the image-quality metrics (PSNR, MS-SSIM, LPIPS) are computed on rendered views: we render both the reference and the decompressed asset from $36$ viewpoints placed on a sphere using a Fibonacci lattice, at a resolution of $512\times512$ with a perspective camera and Phong shading, and average each metric over all views and all objects. The geometric distortion metrics (Chamfer Distance, EMD) are computed on $2048$ points sampled from each surface, with both point sets normalized into the unit cube $[0,1]^3$.

In addition to file-relative compression ratios, we report standardized rates in~\Cref{tab:standardized_rates}. Mesh rates are bits per input vertex and point-cloud rates are bits per input point; lower values indicate smaller compressed representations for a fixed input element count.

\begin{table*}[tb]
    \centering
    \pinkall
    \small
    \caption{\textbf{Standardized Rate Measures.} Bits per input vertex for mesh compression and bits per input point for point-cloud compression.}
    \begin{tabular}{lrlr}
        \toprule
        \rowcolor{blue!15}Mesh method & Bits/vertex $\downarrow$ & Point-cloud method & Bits/point $\downarrow$\\
        \midrule
        Draco$^\ast$~\citep{galligan2018google} & 141.06 & Draco$^\ddagger$~\citep{galligan2018google} & 29.92\\
        Draco$^\dagger$~\citep{galligan2018google} & 145.61 & Draco$^\mathsection$~\citep{galligan2018google} & 20.88\\
        Draco$^\ddagger$~\citep{galligan2018google} & 147.13 & G-PCC~\citep{schwarz2018emerging} & 12.60\\
        Draco$^\mathsection$~\citep{galligan2018google} & 157.75 & V-PCC~\citep{liu2019comprehensive} & \secondbest{9.32}\\
        Corto~\citep{corto} & 21.24 & SparsePCGC~\citep{9968173} & 13.89\\
        Neural Subd.~\citep{liu2020neural} & 86.46 & \acronym\ (LION, $d_C=512$) & \best{8.00}\\
        DeepSDF~\citep{park2019deepsdf} & 7.58 & &\\
        NGF~\citep{10.1145/3641519.3657399} & 22.75 & &\\
        3DShape2VecSet~\citep{zhang20233dshape2vecset} & \secondbest{2.86} & &\\
        \acronym\ (InstantMesh) & \best{0.46} & &\\
        \bottomrule
    \end{tabular}
    \label{tab:standardized_rates}
\end{table*}

\subsection{Baselines}

We compare \acronym\ against established compression methods for each representation, described below. For radiance fields we compare against codecs applicable to arbitrary (rgb$\sigma$) grids, as most NeRF and 3D Gaussian Splat (3DGS)-specific methods do not apply to our setting.

\textbf{Draco.} Draco~\citep{galligan2018google} is a non-learned codec that we apply to both meshes and point clouds. We always use a compression level of $10$ and report several quantization settings, given as the number of bits for the position / texture-coordinate / normal / generic attributes: $^\mathsection$ uses $14/14/14/14$, $^\ddagger$ uses $11/10/8/8$, $^\dagger$ uses $7/7/7/7$, and $^\ast$ uses $4/4/4/4$. For meshes we report all four settings; for point clouds we report the $^\ddagger$ and $^\mathsection$ settings.

\textbf{Corto.} We compress meshes with Corto~\citep{corto} using $4$-bit quantization for both the normal and texture-coordinate attributes, with the texture image stored separately. The compressed size is read directly from the encoder output.

\textbf{Neural Subdivision.} Neural Subdivision~\citep{liu2020neural} learns to refine a coarse mesh. We train the per-object network for $700$ epochs (learning rate $2\times10^{-3}$) and apply $2$ subdivision levels at inference; its three sub-networks (initialization, edge, and vertex) each use two hidden layers of width $32$.

\textbf{DeepSDF.} DeepSDF~\citep{park2019deepsdf} represents a shape as a latent code decoded by an $8$-layer, width-$512$ MLP (the code is injected at the middle layer, with dropout $0.2$), trained with code regularization for $2001$ epochs. The compressed representation is the per-shape $256$-dimensional FP-32 latent code ($1$\,KB).

\textbf{Neural Geometry Fields (NGF).} NGF~\citep{10.1145/3641519.3657399} fits a neural displacement field over a coarse base mesh. We use $1000$ patches with $20$-dimensional feature vectors and optimize each object over base resolutions $\{4,8,12,16\}$ for $100$ iterations each (Adam, learning rate $10^{-3}$); the compressed size is the serialized feature stream.

\textbf{3DShape2VecSet.} We use the pre-trained shape autoencoder of 3DShape2VecSet~\citep{zhang20233dshape2vecset} (the \texttt{kl\_d512\_m512\_l8} variant) as a learned mesh-compression baseline. To compress a mesh, we sample $2048$ points on its surface and encode them into a set of $512$ latent vectors of dimension $8$; this $512\times8$ latent is the compressed geometry. To decompress, we evaluate the decoder's occupancy field on a $128^3$ grid and extract the surface using Marching Cubes.

\textbf{G-PCC.} G-PCC~\citep{schwarz2018emerging} is the MPEG geometry-based point-cloud coding standard, which we evaluate on point clouds using the TMC13 reference software.

\textbf{V-PCC.} V-PCC~\citep{liu2019comprehensive} is the MPEG video-based point-cloud coding standard, evaluated with the TMC2 reference software. We encode with a 3D coordinate bit depth of $10$ and geometry QP $32$, a single $8$-bit colour attribute at QP $28$, occupancy-map precision $4$, and a maximum depth of $12$.

\pink{\textbf{SparsePCGC.} SparsePCGC~\citep{9968173} is a multiscale sparse-convolution codec. For point clouds, we evaluate the released checkpoints and report the resulting operating point in~\Cref{tab:pcresults,tab:standardized_rates}. For radiance-field (rgb$\sigma$) grids, we use its $8$-stage model with kernel size $5$ and obtain different rate points by varying the position quantization scale.}

\textbf{VQRF.} VQRF~\citep{10204677} compresses radiance-field grids through voxel pruning and vector quantization. We use a codebook of $4096$ entries together with importance-based voxel pruning (inclusion and pruning thresholds of $0.6$ and $0.999$).

\subsection{Collection of Additional Meshes for Evaluation}
\label{sec:collection}

Since \acronym\ is designed as a compression-decompression framework that should not require per-scene retraining, the generalization capability of our mapping networks to unseen 3D models is very important. Thus, we also collected a very diverse dataset of 500 high-quality 3D meshes from the following sources:

\begin{itemize}
    \item OpenLRM Demo: \url{https://huggingface.co/spaces/zxhezexin/OpenLRM}
    \item InstantMesh Demo: \url{https://huggingface.co/spaces/TencentARC/InstantMesh}
    \item Hunyuan3D-2 Demo: \url{https://huggingface.co/spaces/tencent/Hunyuan3D-2}
    \item TRELLIS Demo: \url{https://huggingface.co/spaces/JeffreyXiang/TRELLIS}
    \item SPAR3D Demo: \url{https://huggingface.co/spaces/stabilityai/stable-point-aware-3d}
    \item SORA-3D Demo: \url{https://huggingface.co/spaces/ginipick/SORA-3D}
    \item A subset of ABO~\citep{collins2022abo} composed of 384 meshes
\end{itemize}

\subsection{Collection of Additional Radiance Fields for Evaluation}
\label{sec:rfcollection}

Since \acronym\ is designed as a compression-decompression framework that should not require per-scene retraining, the generalization capability of our mapping networks to unseen 3D models is very important. Thus, we also report results on the unseen ``ai'' subset of the NeRF-MAE dataset~\citep{irshad2024nerf}.

\FloatBarrier
\section{Additional Results}
\label{app:results}

\subsection{Additional Results for Mesh Compression}
\label{app:meshresults}

We show the results for mesh compression on our test set, or the $\mathcal{T}$ set, with 3D metrics (Chamfer distance and EMD) as well as wall-clock times for compression and decompression in~\Cref{tab:500results}.

\begin{table*}[t]
    \centering
    \caption{\textbf{Mesh Compression.} Quantitative comparison of \acronym\ with state-of-the-art 3D mesh compression methods, including geometry-based metrics Chamfer Distance and EMD. We also report the compression and decompression wall-clock times. All quality metrics are reported on our Objaverse test set, denoted $\mathcal{T}$.}
    \begin{adjustbox}{width=\textwidth}
    \begin{tabular}{lrrrrrrrr}
        \toprule
        \rowcolor{blue!15}Method & CR ($\times$) \textbf{\scriptsize\textcolor{gray}{(MB)}} $\uparrow$ & Compress (ms) $\downarrow$ & Decompress (ms) $\downarrow$ & $\mathcal{T}$-PSNR $\uparrow$ & $\mathcal{T}$-MS-SSIM $\uparrow$ & $\mathcal{T}$-LPIPS $\downarrow$ & $\mathcal{T}$-Chamfer $\downarrow$ & $\mathcal{T}$-EMD $\downarrow$\\
        \midrule
        Draco$^\ast$~\citep{galligan2018google} & 6.92 \textbf{\scriptsize\textcolor{gray}{(6.43 / 0.93)}} & 70.30 & 29.97 & 15.45 {\textbf{\scriptsize\textcolor{gray}{($\pm$1.27)}}} & 0.7468 {\textbf{\scriptsize\textcolor{gray}{($\pm$0.09)}}} & 0.2437 {\textbf{\scriptsize\textcolor{gray}{($\pm$0.09)}}} & 1.2746 & 0.0544\\
        Draco$^\dagger$~\citep{galligan2018google} & 6.70 \textbf{\scriptsize\textcolor{gray}{(6.43 / 0.96)}} & 69.05 & 28.33 & 23.33 {\textbf{\scriptsize\textcolor{gray}{($\pm$1.32)}}} & 0.9576 {\textbf{\scriptsize\textcolor{gray}{($\pm$0.03)}}} & 0.1039 {\textbf{\scriptsize\textcolor{gray}{($\pm$0.06)}}} & 0.8739 & 0.0385\\
        Draco$^\ddagger$~\citep{galligan2018google} & 6.61 \textbf{\scriptsize\textcolor{gray}{(6.43 / 0.97)}} & 68.15 & 27.46 & \secondbest{38.91} {\textbf{\scriptsize\textcolor{gray}{($\pm$1.30)}}} & \secondbest{0.9992} {\textbf{\scriptsize\textcolor{gray}{($\pm$0.00)}}} & \secondbest{0.0045} {\textbf{\scriptsize\textcolor{gray}{($\pm$0.00)}}} & 0.4214 & 0.0108\\
        Draco$^\mathsection$~\citep{galligan2018google} & 6.20 \textbf{\scriptsize\textcolor{gray}{(6.43 / 1.04)}} & 66.30 & 25.81 & \best{48.55} {\textbf{\scriptsize\textcolor{gray}{($\pm$1.54)}}} & \best{1.0000} {\textbf{\scriptsize\textcolor{gray}{($\pm$0.00)}}} & \best{0.0004} {\textbf{\scriptsize\textcolor{gray}{($\pm$0.00)}}} & \best{0.2060} & \best{0.0044}\\
        Corto~\citep{corto} & 45.93 \textbf{\scriptsize\textcolor{gray}{(6.43 / 0.14)}} & \secondbest{50.52} & \secondbest{8.20} & 20.92 {\textbf{\scriptsize\textcolor{gray}{($\pm$2.79)}}} & 0.8619 {\textbf{\scriptsize\textcolor{gray}{($\pm$0.08)}}} & 0.1374 {\textbf{\scriptsize\textcolor{gray}{($\pm$0.08)}}} & 0.9644 & 0.0438\\
        Neural Subd.~\citep{liu2020neural} & 11.28 \textbf{\scriptsize\textcolor{gray}{(6.43 / 0.57)}} & 61104.12 & \best{0.00} & 15.95 {\textbf{\scriptsize\textcolor{gray}{($\pm$2.18)}}} & 0.8525 {\textbf{\scriptsize\textcolor{gray}{($\pm$0.04)}}} & 0.1513 {\textbf{\scriptsize\textcolor{gray}{($\pm$0.05)}}} & 1.1125 & 0.0496\\
        DeepSDF~\citep{park2019deepsdf} & 131.22 \textbf{\scriptsize\textcolor{gray}{(6.43 / 0.05)}} & 887.78 & 578.53 & 8.47 {\textbf{\scriptsize\textcolor{gray}{($\pm$0.23)}}} & 0.7039 {\textbf{\scriptsize\textcolor{gray}{($\pm$0.07)}}} & 0.3704 {\textbf{\scriptsize\textcolor{gray}{($\pm$0.08)}}} & 1.5842 & 0.0617\\
        NGF~\citep{10.1145/3641519.3657399} & 42.87 \textbf{\scriptsize\textcolor{gray}{(6.43 / 0.15)}} & 152637.87 & 507.21 & 35.45 {\textbf{\scriptsize\textcolor{gray}{($\pm$3.02)}}} & 0.9987 {\textbf{\scriptsize\textcolor{gray}{($\pm$0.08)}}} & 0.0054 {\textbf{\scriptsize\textcolor{gray}{($\pm$0.03)}}} & \secondbest{0.3866} & 0.0130\\
        3DShape2VecSet~\citep{zhang20233dshape2vecset} & \secondbest{342.50} \textbf{\scriptsize\textcolor{gray}{(6.43 / 0.0188)}} & \best{4.90} & 373.00 & 21.65 {\textbf{\scriptsize\textcolor{gray}{($\pm$2.22)}}} & 0.9047 {\textbf{\scriptsize\textcolor{gray}{($\pm$0.06)}}} & 0.1582 {\textbf{\scriptsize\textcolor{gray}{($\pm$0.09)}}} & 0.5304 & 0.0101\\
        \acronym\ (InstantMesh) & \best{2187.07} \textbf{\scriptsize\textcolor{gray}{(6.43 / 0.003)}} & 270.24 & 1476.00 & 27.50 {\textbf{\scriptsize\textcolor{gray}{($\pm$3.13)}}} & 0.9796 {\textbf{\scriptsize\textcolor{gray}{($\pm$0.02)}}} & 0.0274 {\textbf{\scriptsize\textcolor{gray}{($\pm$0.02)}}} & 0.3930 & \secondbest{0.0084}\\
        \bottomrule
    \end{tabular}
    \end{adjustbox}
    \label{tab:500results}
\end{table*}

\begin{table*}[t]
    \centering
    \caption{\textbf{Radiance Field Results.} Quantitative comparison of \acronym\ for compressing radiance fields from the ai subset of the NeRF-MAE dataset (\Cref{sec:rfcollection}).}
    \begin{adjustbox}{width=\textwidth}
    \begin{tabular}{lrrrrrr}
        \toprule
        \rowcolor{blue!15}Method & CR ($\times$) {\textbf{\scriptsize\textcolor{gray}{(MB)}}} $\uparrow$ & Compress (ms) $\downarrow$ & Decompress (ms) $\downarrow$ & PSNR $\uparrow$ & MS-SSIM $\uparrow$ & LPIPS $\downarrow$\\
        \midrule
        Ours & 657.89 {\textbf{\scriptsize\textcolor{gray}{(59.21 / 0.09)}}} & 45.63 & 75.80 & 22.40 & 0.8958 & 0.1400\\
        \bottomrule
    \end{tabular}
    \end{adjustbox}
    \label{tab:oodrfresults}
\end{table*}
\subsection{Additional Results for Point Cloud Compression}
\label{app:pcresults}

\begin{wrapfigure}[24]{r}{0.45\textwidth}
    \centering
    \vspace{-\baselineskip}
    \includegraphics[width=\linewidth]{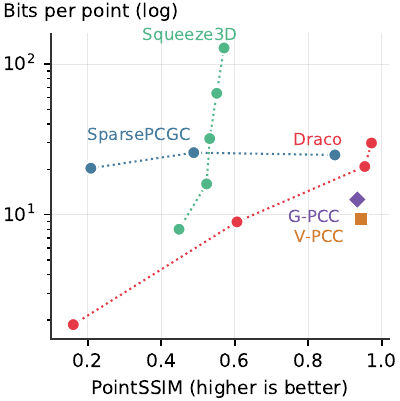}
    \caption{\pink{\textbf{Point-cloud rate--distortion.} PointSSIM (x-axis) versus bits/input point (log y-axis) for \acronym\ and four codecs. Each \acronym\ point uses separately trained mappings.}}
    \label{fig:pc_pareto_sparsepcgc}
\end{wrapfigure}

\pink{We report point-cloud geometry metrics in~\Cref{tab:pcresults_geometry} and rate-distortion behavior in~\Cref{fig:pc_pareto_sparsepcgc}.}

\begin{table*}[t]
    \centering
    \pinkall
    \caption{\textbf{Point Cloud Compression with Geometry Metrics.} Quantitative comparison of \acronym\ with state-of-the-art point cloud compression methods, including geometry-based metrics Chamfer Distance and EMD.}
    \begin{adjustbox}{width=\textwidth}
    \begin{tabular}{lrrrrrrr}
        \toprule
        \rowcolor{blue!15}Method & CR ($\times$) \textbf{\scriptsize\textcolor{gray}{(KB)}} $\uparrow$ & Compress (ms) $\downarrow$ & Decompress (ms) $\downarrow$ & PCQM $\uparrow$ & PointSSIM $\uparrow$ & Chamfer $\downarrow$ & EMD $\downarrow$\\
        \midrule
        Draco$^\ddagger$~\citep{galligan2018google} & 15.64 \textbf{\scriptsize\textcolor{gray}{(117 / 7.48)}} & \secondbest{0.35} & \secondbest{0.34} & \best{3.2875} {\textbf{\scriptsize\textcolor{gray}{($\pm$ 0.13)}}} & \best{0.9722} {\textbf{\scriptsize\textcolor{gray}{($\pm$ 0.11)}}} & \best{0.0284} & \best{0.0061}\\
        Draco$^\mathsection$~\citep{galligan2018google} & \secondbest{22.41} \textbf{\scriptsize\textcolor{gray}{(117 / 5.22)}} & \best{0.25} & \best{0.23} & \secondbest{2.1039} {\textbf{\scriptsize\textcolor{gray}{($\pm$ 0.08)}}} & \secondbest{0.9535} {\textbf{\scriptsize\textcolor{gray}{($\pm$ 0.12)}}} & 0.0419 & 0.0095\\
        \midrule
        \acronym\ (LION) & \best{58.50} \textbf{\scriptsize\textcolor{gray}{(117 / 2.00)}} & 3.85 & 12.74 & 1.8437 {\textbf{\scriptsize\textcolor{gray}{($\pm$ 1.13)}}} & 0.4484 {\textbf{\scriptsize\textcolor{gray}{($\pm$ 0.11)}}} & \secondbest{0.0285} & \secondbest{0.0062}\\
        \bottomrule
    \end{tabular}
    \end{adjustbox}
    \label{tab:pcresults_geometry}
\end{table*}

\subsection{Additional Results for Radiance Field Compression}
\label{app:rfresults}

Our mapping networks are trained on the dataset we derived from NeRF-MAE~\citep{irshad2024nerf}, but we test \acronym\ on only the radiance fields in the \texttt{ai} subset of the dataset, which was collected separately from the other datasets. On our collection of radiance fields, \acronym\ achieves on average only a 4.22 dB reduction in the PSNR as we show in~\Cref{tab:oodrfresults}.

\begin{wrapfigure}[20]{r}{0.6\textwidth}
    \centering
    \vspace{-2\baselineskip}
    \includegraphics[width=\linewidth]{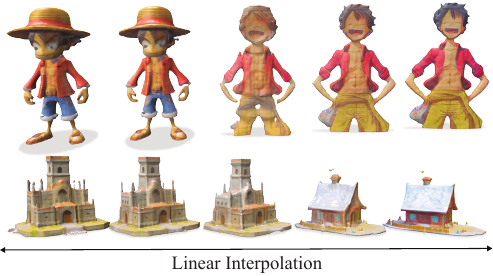}
    \caption{\textbf{Interpolation.} The compressed representations we obtain can also be interpolated. In these examples, we obtain the compressed representation for the leftmost and rightmost meshes and linearly interpolate between them.}
    \label{fig:interp}
\end{wrapfigure}

\subsection{Qualitative Results on Meshes}
\label{app:qualmesh}

\begin{figure*}[t]
    \centering
    \includegraphics[width=\textwidth]{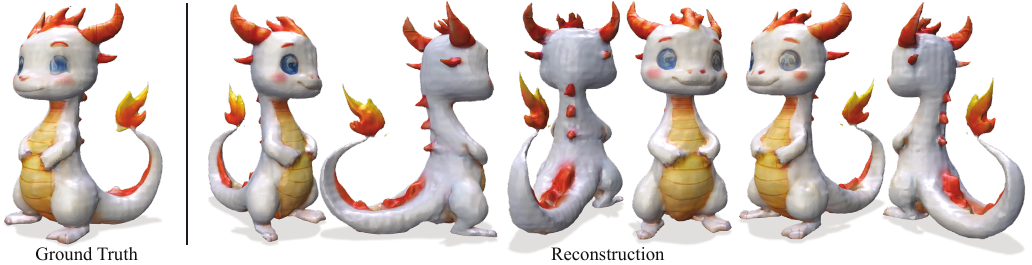}
    \caption{\textbf{Multi-view visualization of compressed and reconstructed meshes.} The consistent appearance across different viewing angles demonstrates that \acronym\ learns correct transformations between latent spaces and produces coherent 3D reconstructions. This confirms that our compressed representation encodes complete 3D information rather than view-dependent features.}
    \label{fig:rotated}
\end{figure*}

\begin{figure}[t]
    \centering
    \includegraphics[width=0.6\linewidth]{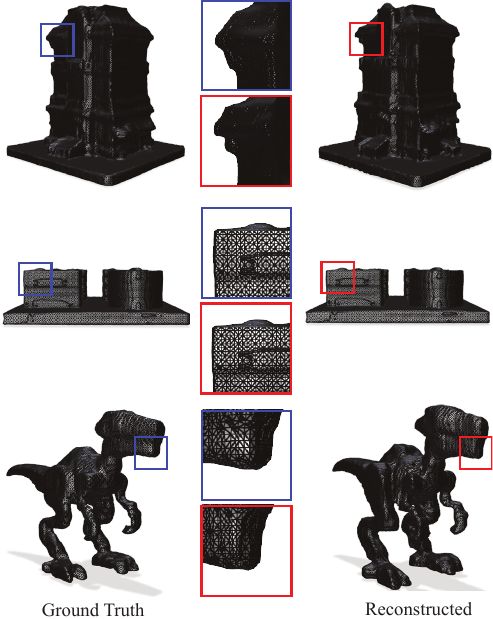}
    \caption{\textbf{\acronym\ preserves geometry details.} We show some meshes compressed with \acronym\ as wireframes. Notice that \acronym\ preserves many fine-grained geometric details.}
    \label{fig:textureless}
\end{figure}

\begin{figure}[tb]
    \centering
    \begin{minipage}[t]{0.62\textwidth}
        \centering
        \includegraphics[width=\linewidth]{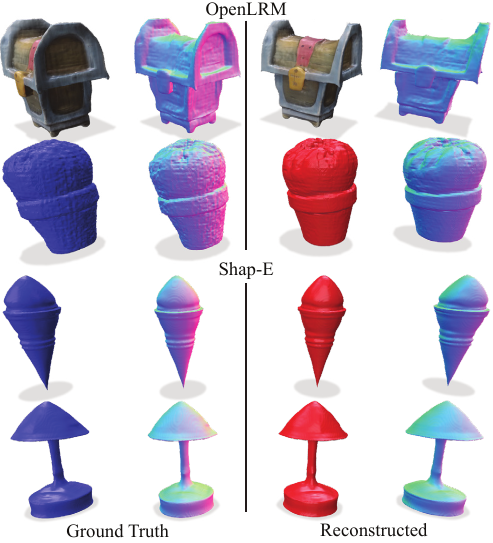}
        \caption{\textbf{Compression results using different 3D generators.} \acronym\ is agnostic to the choice of a 3D generation model. Thus, we show compression results with the 3D generators OpenLRM and Shap-E. We choose 3D meshes that lie within the representational capacity of the chosen 3D generators.}
        \label{fig:decodervariation}
    \end{minipage}
    \hfill
    \begin{minipage}[t]{0.36\textwidth}
        \centering
        \includegraphics[width=\linewidth]{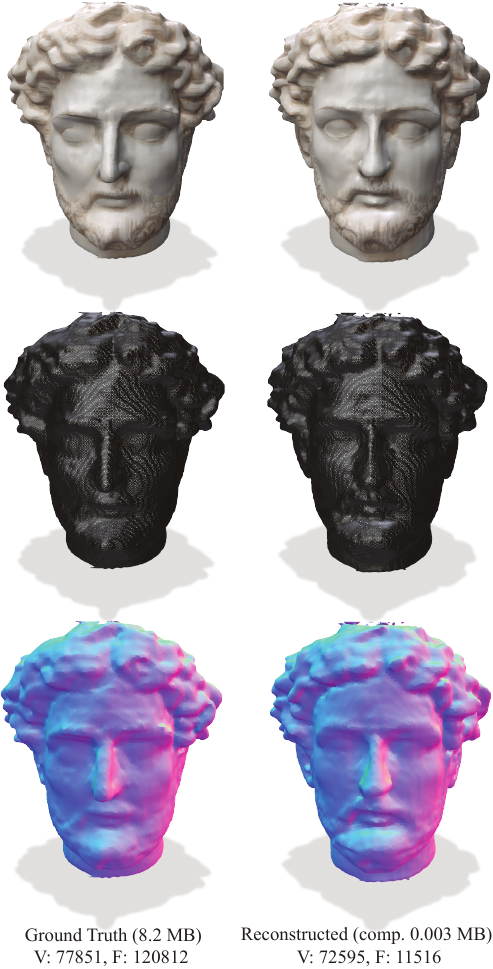}
        \caption{\textbf{Compressing Complex Meshes.} \acronym\ can be used to compress highly complex textured 3D meshes (in this case 77851 vertices and 120812 faces).}
        \label{fig:tradeoff}
    \end{minipage}
\end{figure}

We also notice that our approach works well with high-resolution complex meshes as we show in~\Cref{fig:tradeoff}. In~\Cref{fig:rotated}, we show a mesh we reconstruct with our method from multiple camera angles to demonstrate that our approach learns a correct transformation between the latent spaces and the reconstruction is consistent. We also observe that the compressed representation our method learns can be interpolated as we show in~\Cref{fig:interp}. In~\Cref{fig:decodervariation} we show examples using two different generative models indicating that \acronym\ can be extended to use other generative models for compression. In~\Cref{fig:textureless}, we observe that our method learns to effectively represent intricate geometrical details.

\WFclear
\Needspace{27\baselineskip}


\subsection{Ablations}
\label{app:ablations}
We conduct several ablation studies to better understand the behavior of \acronym\ and analyze the trade-offs between compression ratio, reconstruction quality, and computational efficiency. First, we investigate how varying the size of our compressed representation affects decompression time. In~\Cref{tab:ablations} on a separate test set of 100 point clouds, we show that increasing the compressed size modestly increases compression and decompression times because it changes the size of the mapping networks. Next, we examine how different compression sizes affect reconstruction quality.~\Cref{tab:ablations} shows a correlation between latent dimension and reconstruction quality. We observe substantial improvements in PointSSIM when increasing the compressed size from 1024 to 2048; however, this improvement in quality starts diminishing beyond the compressed size of 2048. We perform an ablation of training with and without Gram loss (\Cref{sec:improve}) in~\Cref{tab:ablationortho}.

\Needspace{5\baselineskip}
\subsection{Failure Cases}
\label{sec:failure_cases}

\noindent\begin{minipage}[t]{0.52\textwidth}
\pink{To better understand the limitations of our approach, we analyze failure cases where \acronym\ struggles to faithfully reconstruct the input geometry.}
\end{minipage}
\par
\WFclear

\begin{figure}[tb]
    \centering
    \includegraphics[width=\textwidth]{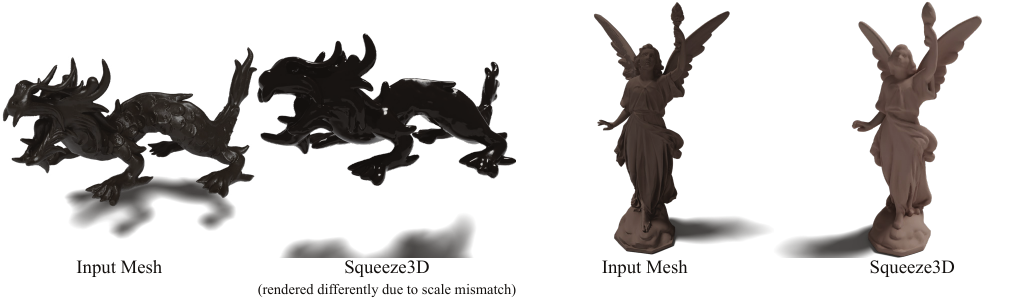}
    \caption{\textbf{Failure Cases.} We show examples where \acronym\ fails to accurately reconstruct the input. (Left) Input mesh with highly intricate details or text. (Right) The reconstruction is smoothed, losing the fine-grained text or surface texture.}
    \label{fig:failure_cases}
\end{figure}

\pink{The primary cause of failure stems from the dependency on the generator's prior distribution. When the input geometry significantly deviates from the distribution of shapes the generator was trained on, the generator may fail to produce the correct details, or the mapping network may fail to find a corresponding code in the generator's latent space that represents the input. We illustrate these failure cases in~\Cref{fig:failure_cases} and~\Cref{tab:failure_metrics}. We primarily observe loss of fine details, where highly intricate structures or text that are not well-represented in the generator's latent space are smoothed out or ignored. This limitation can be due to: (1) The encoder may fail to capture high-frequency details. (2) Even if the encoder captures these details, the generator's latent bottleneck may filter them out.}

\begin{table}[t]
    \centering
    \caption{\textbf{Ablations.} Ablation study on compressed representation size. We analyze how varying the dimensionality of the compressed latent space affects compression time, decompression time, and reconstruction quality (measured by PCQM and PointSSIM).}
    \begin{tabular}{rrrrr}
        \toprule
        \rowcolor{blue!15}Size ($\mathbf{z}_{\text{comp}}$) & Compression (ms) & Decompression (ms) & PCQM $\uparrow$ & PointSSIM $\uparrow$ \\
        \midrule
        512  & \best{3.85} & \best{12.74} & 1.4001 & 0.3640 \\
        1024 & \best{3.85} & \best{12.74} & 1.4047 & 0.3640 \\
        2048 & \secondbest{4.22} & \secondbest{13.39} & 1.3311 & 0.4249 \\
        4096 & 4.85 & 14.04 & \best{1.8437} & \secondbest{0.4318} \\
        8192 & 5.30 & 14.19 & \secondbest{1.5665} & \best{0.4473} \\
        \bottomrule
    \end{tabular}
    \label{tab:ablations}
\end{table}

\begin{table*}[t]
    \centering
    \caption{\textbf{Ablation.} Quantitative comparison of \acronym\ (InstantMesh) with and without Gram-loss.}
    \begin{adjustbox}{width=\textwidth}
    \begin{tabular}{lrrrrrr}
        \toprule
        \rowcolor{blue!15}Method & CR ($\times$) {\textbf{\scriptsize\textcolor{gray}{(MB)}}} $\uparrow$ & Compress (ms) $\downarrow$ & Decompress (ms) $\downarrow$ & PSNR $\uparrow$ & MS-SSIM $\uparrow$ & LPIPS $\downarrow$\\
        \midrule
        \acronym\ (InstantMesh) & \best{2187.07} {\textbf{\scriptsize\textcolor{gray}{(6.43 / 0.003)}}} & \best{270.24} & \best{1476.00} & \best{27.50} {\textbf{\scriptsize\textcolor{gray}{($\pm$3.13)}}} & \best{0.9796} {\textbf{\scriptsize\textcolor{gray}{($\pm$0.02)}}} & \best{0.0274} {\textbf{\scriptsize\textcolor{gray}{($\pm$0.02)}}}\\
        w/o Gram-loss & \best{2187.07} {\textbf{\scriptsize\textcolor{gray}{(6.43 / 0.003)}}} & \best{270.24} & \best{1476.00} & \secondbest{24.20} {\textbf{\scriptsize\textcolor{gray}{($\pm$3.50)}}} & \secondbest{0.9205} {\textbf{\scriptsize\textcolor{gray}{($\pm$0.04)}}} & \secondbest{0.1321} {\textbf{\scriptsize\textcolor{gray}{($\pm$0.035)}}}\\
        \bottomrule
    \end{tabular}
    \end{adjustbox}
    \label{tab:ablationortho}
\end{table*}

\begin{table}[t]
    \centering
    \pinkall
    \small
    \caption{\textbf{Quantitative failure-case analysis.} Image metrics are means over 36 geometry-only views; normalized Chamfer distance and approximate 3D EMD use 2,048 surface samples. Both meshes are independently normalized to a unit box without rigid ICP alignment.}
    \begin{tabular}{lrrrrr}
        \toprule
        \rowcolor{blue!15}Case & PSNR ($\uparrow$) & MS-SSIM ($\uparrow$) & LPIPS ($\downarrow$) & Chamfer ($\downarrow$) & EMD ($\downarrow$) \\
        \midrule
        XYZRGB Dragon & 21.13 & 0.9315 & 0.0611 & 0.05154 & 0.12827 \\
        Lucy & 19.47 & 0.8979 & 0.1043 & 0.19425 & 0.35322 \\
        \bottomrule
    \end{tabular}
    \label{tab:failure_metrics}
\end{table}

\begin{table*}[tb]
    \centering
    \pinkall
    \caption{\textbf{Failure attribution.} For the encoder-side test, we decode the MeshAnything representation using one greedy and three stochastic native autoregressive decodes and select the lowest-Chamfer result. For the generator-side test, we condition frozen InstantMesh directly on six target views, bypass Zero123Plus, evaluate two fixed seeds, and select the lower-Chamfer result.}
    \begin{adjustbox}{width=\textwidth}
    \begin{tabular}{ll*{5}{r}}
        \toprule
        \rowcolor{blue!15}Case & Reconstruction path & PSNR ($\uparrow$) & MS-SSIM ($\uparrow$) & LPIPS ($\downarrow$) & Chamfer ($\downarrow$) & EMD ($\downarrow$) \\
        \midrule
        XYZRGB Dragon & Native MeshAnything & 22.93 & 0.9600 & 0.0321 & 0.01670 & 0.01936 \\
        XYZRGB Dragon & Direct InstantMesh & 18.21 & 0.8795 & 0.1093 & 0.15039 & 0.28745 \\
        Lucy & Native MeshAnything & 24.19 & 0.9607 & 0.0422 & 0.02736 & 0.07611 \\
        Lucy & Direct InstantMesh & 21.25 & 0.9252 & 0.0743 & 0.05875 & 0.07850 \\
        \bottomrule
    \end{tabular}
    \end{adjustbox}
    \label{tab:failure_attribution}
\end{table*}

\pink{To identify which component limits these examples, we decode the MeshAnything representation using its native decoder and also reconstruct each object directly with InstantMesh, bypassing our mapping networks. The MeshAnything result is selected from one greedy and three stochastic decodes; the InstantMesh result is selected from two fixed-seed reconstructions conditioned directly on the six target views without Zero123Plus. As shown in~\Cref{tab:failure_attribution}, the native MeshAnything reconstruction substantially improves both failures. Direct InstantMesh reconstruction remains worse than native MeshAnything for both objects. These results indicate that the generator prior is the dominant quality ceiling for these cases.}

\Needspace{5\baselineskip}
\subsection{Robustness to Input Noise}
\label{sec:noise_robustness}

\pink{We perform a sensitivity analysis by adding zero-mean i.i.d.\ Gaussian noise to the vertices of every mesh in the $\mathcal{T}$-set and rerunning the complete encoder--mapping--generator pipeline. Here $\sigma$ is the per-axis standard deviation expressed as a fraction of each clean mesh's bounding-box diagonal. For each mesh, we use the same noise directions, sampled faces, and barycentric coordinates at every noise level. \Cref{tab:noise_robustness} indicates modest changes at the smallest perturbation and increasingly strong degradation at larger noise levels.}

\begin{table}[t]
    \centering
    \pinkall
    \caption{\textbf{Input-noise sensitivity.}}
    \begin{tabular}{lrrrrr}
        \toprule
        \rowcolor{blue!15}$\sigma$ / bbox diagonal & PSNR ($\uparrow$) & MS-SSIM ($\uparrow$) & LPIPS ($\downarrow$) & Chamfer ($\downarrow$) & EMD ($\downarrow$) \\
        \midrule
        0 (clean) & \best{27.50} & \best{0.9796} & \best{0.0274} & \best{0.3930} & \best{0.0084} \\
        0.005 & \secondbest{26.49} & \secondbest{0.9767} & \secondbest{0.0329} & \secondbest{0.4517} & \secondbest{0.0174} \\
        0.010 & 24.19 & 0.9663 & 0.0533 & 0.6257 & 0.0204 \\
        0.025 & 20.12 & 0.9280 & 0.1369 & 1.1186 & 0.0305 \\
        0.050 & 19.07 & 0.9125 & 0.1916 & 1.9091 & 0.0954 \\
        0.100 & 18.54 & 0.9046 & 0.2140 & 2.0378 & 0.1107 \\        
        \bottomrule
    \end{tabular}
    \label{tab:noise_robustness}
\end{table}

\pink{We analogously evaluate radiance-field sensitivity by adding zero-mean i.i.d.\ Gaussian noise to every (rgb$\sigma$) grid in the radiance-field $\mathcal{T}$-set. Here $\sigma$ is expressed as a fraction of each input's dynamic range. \Cref{tab:rf_noise_robustness} shows gradual degradation as the perturbation increases.}

\begin{table}[t]
    \centering
    \pinkall
    \caption{\textbf{Radiance-field input-noise sensitivity.} Gaussian noise scale is expressed relative to the input dynamic range.}
    \begin{tabular}{lrrr}
        \toprule
        \rowcolor{blue!15}$\sigma$ / input dynamic range & PSNR ($\uparrow$) & MS-SSIM ($\uparrow$) & LPIPS ($\downarrow$)\\
        \midrule
        0 (clean) & \best{26.62} & \best{0.9533} & \best{0.0743}\\
        0.005 & \secondbest{26.55} & \secondbest{0.9528} & \secondbest{0.0752}\\
        0.010 & 26.43 & 0.9520 & 0.0768\\
        0.025 & 26.10 & 0.9498 & 0.0805\\
        0.050 & 25.62 & 0.9460 & 0.0869\\
        0.100 & 24.90 & 0.9395 & 0.0980\\
        \bottomrule
    \end{tabular}
    \label{tab:rf_noise_robustness}
\end{table}

\Needspace{4\baselineskip}
\subsection{Decoder Amortization}
\label{app:amortization}

\pink{We define decoder amortization as the number of objects after which storing the frozen decoder and reverse mapping network once plus the compressed latent for each object uses no more storage than the original objects. For shared decompression storage $S_{\mathrm{shared}}$, average original size $S_{\mathrm{orig}}$, and per-object latent size $S_{\mathrm{latent}}$, the break-even count is $\left\lceil S_{\mathrm{shared}}/(S_{\mathrm{orig}}-S_{\mathrm{latent}})\right\rceil$. \Cref{tab:amortize_thresholds} reports the thresholds.}

\begin{table}[t]
    \centering
    \pinkall
    \small
    \caption{\textbf{Decompression-side amortization thresholds.} Exact break-even calculations including both the frozen decoder and reverse mapping network.}
    \begin{tabular}{lrrrr}
        \toprule
        \rowcolor{blue!15}Decoder & Shared storage & Original/object & Latent/object & Break-even objects \\
        \midrule
        Shap-E      & 2.89\,GB & 6.43\,MB & 4.00\,KB & 449 \\
        InstantMesh & 3.43\,GB & 6.43\,MB & 3.01\,KB & 534 \\
        OpenLRM     & 2.79\,GB & 6.43\,MB & 4.00\,KB & 435 \\
        LION ($d_C=512$) & 312.25\,MB & 117\,KB & 2.00\,KB & 2,716 \\
        NeRF-MAE    & 478.92\,MB & 58.07\,MB & 93.75\,KB & 9 \\
        \bottomrule
    \end{tabular}
    \label{tab:amortize_thresholds}
\end{table}

\Needspace{10\baselineskip}
\section{Additional Related Works}

While we are interested in learning-based methods for compression, we include
classical compression techniques, which are only tangentially related to our work, for completeness.
These include approaches that reorder the structure of triangles and faces in the mesh to enable compressed encodings of elements based on their local structure and perform quantization~\citep{10.1145/218380.218391, 10.1145/274363.274365, 764870, 29920bfb85ab46c2a4aff778703182ca, galligan2018google}. Most of these techniques are lossless and thus achieve high fidelity; however, they are inherently limited in their ability to significantly reduce file sizes since they preserve all details of the mesh representation.
To overcome the limitations of lossless compression, lossy techniques have emerged as popular alternatives. Geometry simplification methods aim to reduce the number of polygons in a mesh while retaining as much of the original structure as possible~\citep{https://doi.org/10.1111/cgf.13932, 10.1145/258734.258849, surazhsky2003explicit, szymczak2002piecewise}. Extensions of these methods have incorporated surface intrinsic properties, such as the mesh Laplacian, as a basis for simplification~\citep{https://doi.org/10.1111/cgf.13932} or error metrics~\citep{10.1145/258734.258849, cohen1998appearance} and coupled this with entropy coding~\citep{corto}. Recently, there have also been approaches that couple these with learned models~\citep{potamias2022neural}.

Traditional subdivision algorithms~\citep{zorin1996interpolating, catmull1998recursively, loop1987smooth} refine coarse meshes by splitting polygonal faces into finer elements, often paired with displacement mapping~\citep{hoppe1994piecewise} to enhance detail. However, these methods rely on hard-coded priors and fixed polynomial interpolations, which can overly smooth the reconstructed geometry and fail to capture intricate details. Neural approaches~\citep{hertz2020deep, liu2020neural} address these limitations by embedding geometric information into learnable parameters.

\section{Results Library}

We present additional results in~\Cref{rl:1,rl:2,rl:3,rl:4}. More results are available in the Supplementary Material.

\foreach \n in {1,...,4}{\begin{figure*}[tb]
    \centering
    \includegraphics[width=\textwidth]{gallery/mesh_gallery_\n.pdf}
    \caption{Results Library.}
    \label{rl:\n}
  \end{figure*}
}

\end{document}